\documentclass[fleqn,preprint]{aa}
\usepackage{mathtools}
\usepackage{txfonts}
\usepackage{graphicx}
\usepackage{epstopdf}
\usepackage{epsfig}
\usepackage{pgf}
\usepackage{amsmath}
\usepackage{physics}

\usepackage{soul}
\definecolor{lightgreen}{HTML}{B7F774}
\sethlcolor{yellow}

\newcommand{\llave}[1]{\left\{ #1 \right\}}

\newcommand{\paren}[1]{\left( #1 \right)}
\newcommand{\corch}[1]{\left[ #1 \right]}

\begin{document}

\title{Derivation of the physical parameters of the jet in S5\,0836$+$710 from stability analysis}

\author{L.~Vega-Garc\'ia\inst{1}%\thanks{email: lauvegar@mpifr-bonn.mpg.de}
        \and
        M.~Perucho\inst{2,3}
        \and
        A.~P.~Lobanov\inst{1,4}
  }

    \institute{Max-Planck-Institut f\"ur Radioastronomie,
              Auf dem H\"ugel 69, D-53121 Bonn, Germany
    \and
    Departament d'Astronomia i Astrof\'isica, Universitat de Val\`encia, C/ Dr. Moliner, 50, E-46100 Burjassot, Val\`encia, Spain 
    \and
    Observatori Astron\`omic, Universitat de Val\`encia, C/ Catedr\`atic Beltr\'an 2, E-46091 Paterna , Val\`encia, Spain
     \and
    Institut f\"ur Experimentalphysik, Universit\"at Hamburg, 
    Luruper Chaussee 149, D-22761 Hamburg, Germany
    }

\authorrunning{Vega-Garc\'ia, Perucho \& Lobanov}
\titlerunning{Instability analysis of S5\,0836$+$710}

  \date{}
 
  \abstract
  % context heading (optional), leave it empty if necessary  
  {A number of extragalactic jets show periodic structures at different scales that can be associated with growing instabilities. The wavelengths of the developing instability modes and their ratios depend on the flow parameters, so the study of those structures can shed light on jet physics at the scales involved.
    }
  {In this work, we use the fits to the jet ridgeline obtained from different observations of S5\,B0836$+$710 and apply stability analysis of relativistic, sheared flows to derive an estimate of the physical parameters of the jet. 
  }
  {Based on the assumption that the observed structures are generated by growing Kelvin-Helmholtz (KH) instability modes, we have run numerical calculations of stability of a relativistic, sheared jet over a range of different jet parameters. We have spanned several orders of magnitude in jet-to-ambient medium density ratio, and jet internal energy, and checked different values of the Lorentz factor and shear layer width. This represents an independent method to obtain estimates of the physical parameters of a jet.}
  {By comparing the fastest growing wavelengths of each relevant mode given by the calculations with the observed wavelengths reported in the literature, we have derived independent estimates of the jet Lorentz factor, specific internal energy, jet-to-ambient medium density ratio and Mach number. We obtain a jet Lorentz factor $\gamma \simeq 12$, specific internal energy of $\varepsilon \simeq 10^{-2}\,c^2$, jet-to-ambient medium density ratio of $\eta\approx 10^{-3}$, and an internal (classical) jet Mach number of $M_\mathrm{j}\approx 12$. We also find that the wavelength ratios are better recovered by a transversal structure with a width of $\simeq 10\,\%$ of the jet radius.}
   {This method represents a powerful tool to derive the jet parameters in all jets showing helical patterns with different wavelengths.}


   \maketitle
%
%________________________________________________________________

\section{Introduction}
 
Extragalactic jets of relativistic plasma are formed in the immediate vicinity of supermassive black holes (SMBH) residing in the nuclei of active galaxies \citep[AGN,][]{blandford+1977}. The jets not only carry energy from this region to hundreds of kiloparsecs, but also condition galaxy evolution since the moment they are initially triggered \citep[see, e.g.,][]{mcnamara+2012}. Therefore, determination of their physical parameters can allow us to understand the precise processes governing them and understand their impact on the ambient medium in more detail. 

Radio images of many extragalactic jets from AGN show helical patterns on parsec and kiloparsec scales \citep[see, e.g.,][]{lister+2013}. 
In different works, these patterns have been associated either with Alfv\'en \citep{cohen2015} or with pressure waves \citep{perucho+2012a}. 
A periodicity in the direction of ejection of the flow due to jet precession is probably the origin of the oscillations.
Precession can arise from the gravitational effect
of the accretion disk on the compact object in the center of the AGN or in a
binary-black-hole \citep[see, e.g.,][]{Lister2003,Stirling2003,Lobanov2005,Bach2006,Savolainen2006}. Once triggered, the induced oscillation can couple to 
instability modes of the Kelvin-Helmholtz instability \citep[KH,][]{perucho+2005,Mizuno2007,Perucho2010} or current driven instability
\citep[CD,][]{Mizuno2009,McKinney2009,Mignone2010,Mizuno2011} and grow in amplitude with time/distance as the wave is advected downstream.

The radio source S5\,0836$+$710 (4C\,$+$71.07; J0841$+$7053) is a powerful
low-polarization quasar (LPQ) located at a redshift of 2.17
\citep{osmer+1994}, which corresponds to the luminosity distance of
16.9\,Gpc and a linear scale of 8.4\,pc/mas,
assuming the standard $\Lambda$CDM cosmology ($H_0 =  73$ km s$^{-1}$ Mpc$^{-1}$, $\Omega_{\mathrm{m}}$ =   0.27, $\Omega_{\mathrm{\Lambda}}$ =   0.73) \citep{planck2015}. 
It has a long and one-sided jet at parsec and kiloparsec scales.
%, which extends to a 
%deprojected distance of $\sim25$\,kpc .
Large-scale radio emission was revealed by the VLA\footnote{Karl
  G. Jansky Very Large Array of the National Radio Astronomy
  Observatory, Socorro, NM, USA} and MERLIN\footnote{Multi-Element
  Radio Linked Interferometer Network of the Jodrell Bank Observatory,
  UK} at distances larger than 1 arcsecond
\citep{hummel+1992,perucho+2012b}. The jet shows apparent speeds of up to $21\,c$ at $15$~GHz, with a mean of $17\,c$ ($\gamma_j \simeq 17$), as reported
by \cite{lister+2013}. The estimated jet viewing angle is $\theta
\approx 3.2 ^\circ$ \citep{otterbein+1998}, who also gave a Lorentz factor of $\gamma_j = 12$ from
jet kinematics at 8~GHz.

The source morphology suggests the existence of plasma instability.
\cite{krichbaum+1990} observed the source with VLBI\footnote{Very Long Baseline Interferometry} at 326\,MHz and 5\,GHz and
recognized several kinks in the flow that can be associated with the
growth of instability. From high-resolution images obtained with VLBI
Space Observatory Program \citep[VSOP;][]{hirabayashi+2000} at 1.6\,GHz and
5\,GHz, \cite{lobanov+1998} suggested the presence of Kelvin-Helmholtz (KH) instability in the
flow, based on the observed morphology of the jet brightness ridgeline. The oscillations in the ridgeline were
identified with helical and elliptical surface modes of the
instability. \cite{perucho+2012a} provided further evidence for the interpretation of these oscillations as caused by KH instability modes. Furthermore, observations with MERLIN show an emission gap between
0.2\,$\arcsec$ and 1.0\,$\arcsec$ and a large-scale, apparently decollimated structure beyond 1~$\arcsec$. This was
explained as the expansion and disruption of the jet due to a helical instability mode \citep{perucho+2012b}.
%The presence of shocks has been suggested in the jet on scales up to
%$\sim 0.5\,$kpc, revealed by regions of flatter spectral indices separated by $\simeq 5~{\rm mas}$
%\citep{lobanov+2006}.

In this paper, we use the results obtained from different observational campaigns on this source \citep[see][]{lobanov+2006,perucho+2012a}, complemented by the Global VLBI array observations performed on October 24th, 2013, (L-Band), as part of {\em RadioAstron} space VLBI observations of S5\,0836+710 (\citealt{VegaGarcia}, Vega-Garc\'ia et al., submitted). We apply a numerical solver of the stability equation for sheared, relativistic flows \citep{perucho+2007a,perucho+2007b} to derive an estimate of the physical parameters in the jet. In contrast with \citet{perucho+2007b}, where the authors fixed the physical parameters of the jet to those obtained via previous estimates \citep{lobanov+1998,lobanov+2006} and studied the possibility of the presence of a shear-layer separating the jet and its environment, here we provide an independent estimate of all the relevant parameters without any initial restriction. The wavelengths corresponding to the fastest growing modes derived from the calculations are compared to the observed wavelengths in the jet, which allows us to provide a set of jet parameters. 

The paper is structured as follows. In Section~\ref{sc:ridgeLineOscillations} we summarize the observational results that have provided fits to the intrinsic wavelengths of the jet ridgeline. Section~\ref{sc:method} includes an explanation of the method that we have used. The results obtained are presented in Section~\ref{sc:jetparameters}. Finally, the discussion of those results and conclusions of this work are given in Section~\ref{sc:summary}.
            
\section{Ridgeline oscillations}
\label{sc:ridgeLineOscillations}

\begin{figure}[htbp!]
  \centering
  \includegraphics[width=0.38\textwidth]{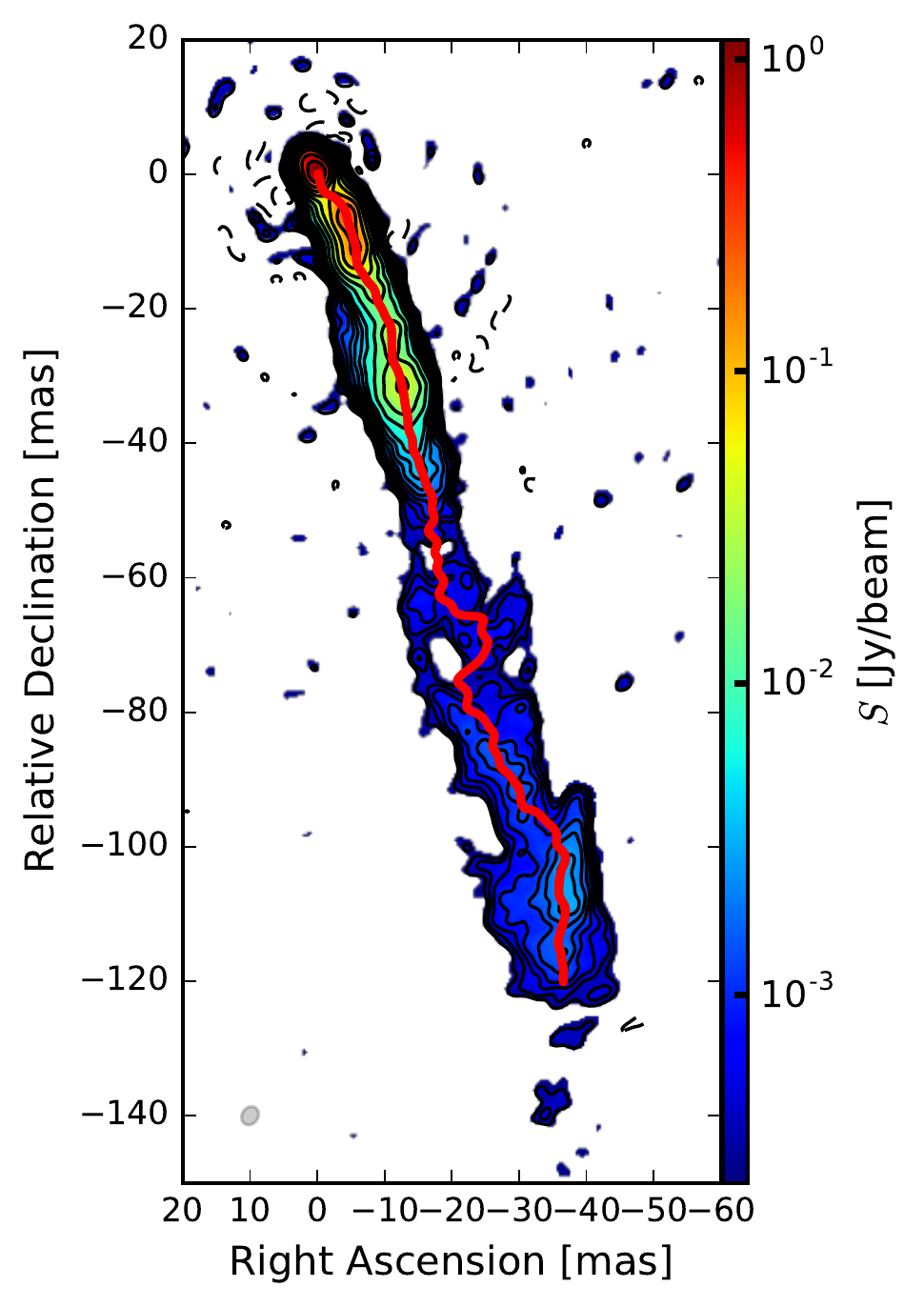}
  \caption{Ridgeline (red) for ground-VLBI observations at 1.6~GHz. 
  The contour levels are drawn at ($-$1, 1, $\sqrt{2}$, 2, ... ) times 0.32~mJy/beam. From \citealt{VegaGarcia}, Vega-Garc\'{\i}a et al. (submitted).}
\label{fg:0836-2-gridge} 
\end{figure}

The ridgeline of the jet is defined as a line connecting the maxima of Gaussian profiles fitted to the jet brightness profiles measured transversally to the jet direction. 
Figure~\ref{fg:0836-2-gridge} shows an example of the ridgeline in the jet in 0836+710 from a VLBI image at 1.6 GHz (\citealt{VegaGarcia}, Vega-Garc\'ia et al., submitted).  At this frequency, the jet in 0836+710 can be traced up to $\approx 130$\,mas (deprojected linear distance of $\approx 10$\,kpc), which enables making detailed studies of a number of instability wavelengths.

\begin{figure}[t!]
\includegraphics[width=0.53\textwidth]{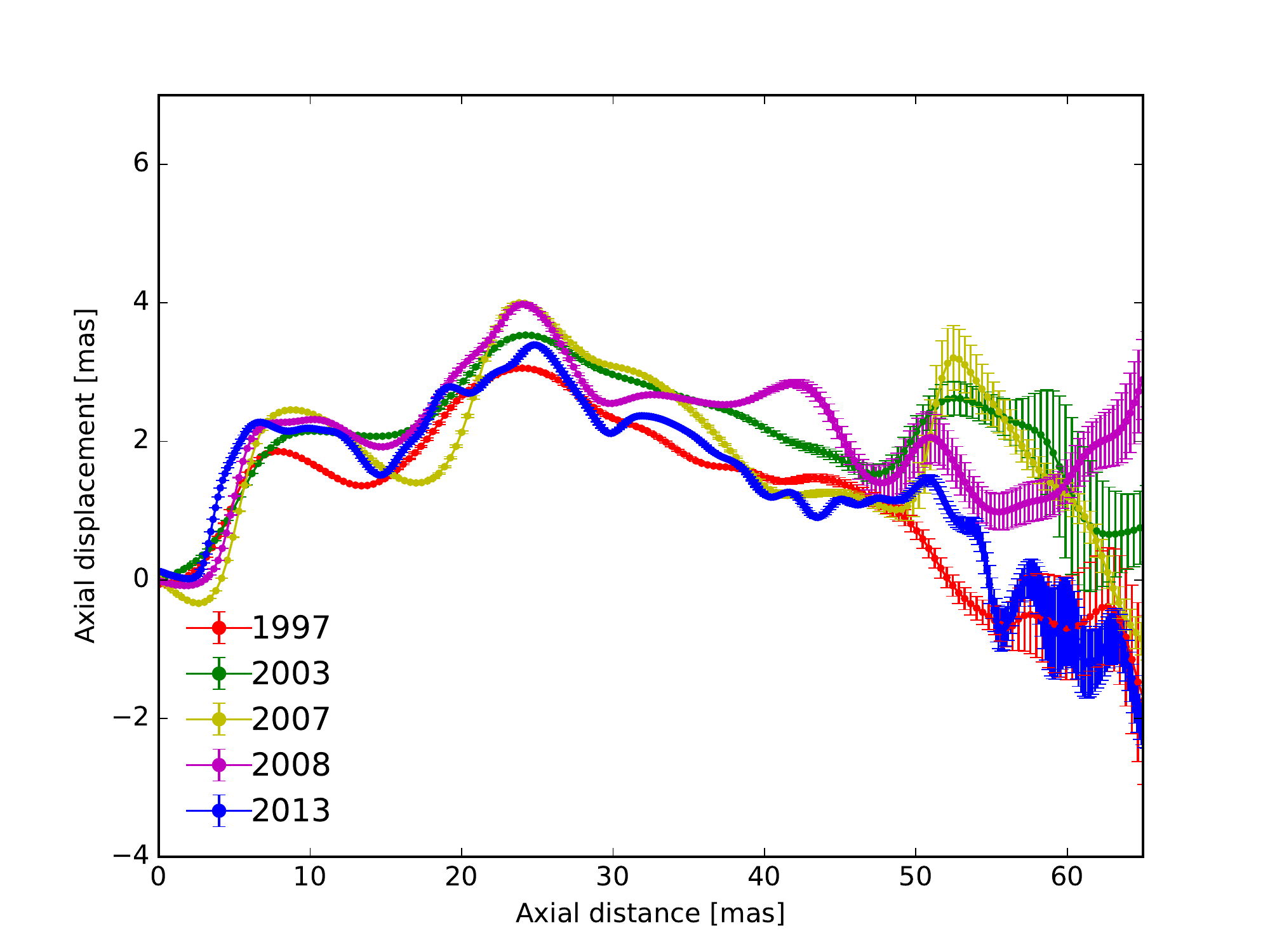}
\caption{Ridgelines at 1.6\,GHz from the observations described in \cite{perucho+2012a} together with those derived from the 2013/2014 Global VLBI observations. The main described structure is similar. The differences are due to the time difference, different data quality and different \emph{uv}-coverages.}
 \label{fig:ridgeComparison}
\end{figure}

 A comparison of the ridgelines obtained from the VLBI images of 0836+710 made at 1.6\,GHz at different epochs shows a consistent picture of a slowly evolving instability pattern in the jet \citep{perucho+2012a}. These ridgelines are shown in Fig.~\ref{fig:ridgeComparison}, tracing the jet up to a distance of 65\,mas. At distances larger than about 60 mas, the jet is substantially resolved in the transverse direction. At these scales the formal ridgeline derived from the brightness profiles is not as reliable for reflecting the plasma instability development, and a more complex description comprising multiple threads inside the flow may be needed \citep[e.g., the one similar to the approach employed for strongly resolved jets in 3C273;][]{lobanov2001}. In addition to this, it is also likely that the plasma instability modes developing in the inner jet would saturate or dissipate at these distances \citep[see][]{perucho+2012a}. Because of these considerations, we limit our analysis of the jet instability in 0836+710 to the structures traced by the jet ridgeline within the inner 65 mas of the flow. The ridgelines offsets are plotted with respect to the overall jet axis assumed to be oriented at a position angle of $-162^\circ$. The small differences observed among the different epochs can be attributed to image noise and small changes in time \citep[][]{perucho+2012a,VegaGarcia}. 
 %, as it was already pointed out by \citet{perucho+2012a}, but for small differences that show up when a prominent double-peaked brightness profile appears between 1 and 4 mas \footnote{These can be due to limb-brightening or to elliptic modes developing along with the helical ones.} or due to differences in the {\em uv}-coverages of the observations, which may trigger small-scale oscillations of the ridgeline. All this information allows to conclude that the large-scale structures have a physical interpretation.
 
 Since the oscillations displace the jet axis from a straight line (as expected for the helical modes of KH instability) and the transverse profiles show hints of double-peaked structure (expected for the elliptical mode of the instability) only in the close vicinity of the core \citet{perucho+2012a}, we consider that helical modes are largely determining the appearance of the ridgelines. The oscillations can thus be modelled as waves developing along the jet. From now on, we assume that these waves correspond to KH instability modes developing in the jet. Therefore, one can take the offset of the ridge from the straight line defining the jet direction as the projected amplitude of the wave, $\Delta\,r(z)$. Taking into account that the jet opening angle is small \citep[$< 1^\circ$,][]{perucho+2012a}, the expected effect of jet expansion on the mode wavelength \citep[$\lambda \propto R$,][]{Hardee1982} can be considered to be small. Then, assuming a constant wave amplitude,\footnote{The instability wave amplitude is expected to grow with distance, but this leaves the wavelengths unaffected so, assuming that the growth lengths are long enough, this represents an acceptable assumption. Otherwise, the growth should be included to improve the quality of the fits.} the offsets can be modelled as:
 
\begin{equation} \label{eq:deltaz}
\Delta\,r (z) = \sum_{i=1}^{N_\mathrm{mod}} a_i \sin(2\,\pi/\lambda_i + \psi_i)\,,
\end{equation}
where $a_i$ is the amplitude, $\psi_i$ the phase, and $\lambda_i$ the wavelength corresponding to the $i$-th mode. The offsets of the ridge on the plane of the sky are not affected by projection effects. However, taking into account that we model a three-dimensional helical structure, there is a distortion on the observed ridgeline produced by travel time effects across the jet cross section. In this case, the distortion is caused by the oscillation in the direction orthogonal to both the jet axis and the direction defined by the observed ridgeline ripples on the plane of the sky. Our approach neglects the role of that distortion because we focus on the longest wavelengths ($\lambda >> R_j$), where these distortions are less relevant than for shorter ones ($\lambda \sim R_j$). 

The $\chi^2$ fits to the ridges can be done in two different ways: either by fitting one mode at a time and substracting it from the original ridgeline before fitting the next  mode, or by fitting all modes simultaneously, starting from an initial set of mode parameters determined in earlier works \citep[e.g.,][]{lobanov+2006,perucho+2012a}. Both approaches give the same results (see, e.g., \citealt{VegaGarcia} for the fits to the ridgline of Global VLBI 1.6 GHz image of the source). For the purposes of this work, we only need the result of this fit, which is reproduced in Table~\ref{tb:modes16}. It is important to stress that the wavelengths obtained by this analysis are similar to those derived in previous works \citep{lobanov+2006,perucho+2012a}, with improved resolution in our case since the observations included the global VLBI array.
%and gives very similar results to those given in previous works, as stated above. The robustness of the result allows us to use them for a further analysis that will end up in the derivation of the jet physical parameters. 

\begin{table}[!tbhp]
\centering
\caption{Results of the fit to the jet ridgeline obtained from the 1.6 GHz image shown in Fig.~\ref{fg:0836-2-gridge}. The fits are shown in Fig.~\ref{fig:modesfit}. Adapted from \citealt{VegaGarcia}, Vega-Garc\'{\i}a et al., submitted.}
%\resizebox{\columnwidth}{!}
{\begin{tabular}{c|ccc}
\hline \hline
Mode &  $\lambda$ [mas]& $a$ [mas] & $\psi$ [$^\circ$] \\
\hline
\hline
 1     & $102 \pm 6$ &  $2.9\pm 0.2$ & $15\pm 8$ \\
 2     & $35\pm 5$ &  $0.21\pm 0.03$ & $195\pm 50$ \\
 3     & $16.5\pm 1.5$ &  $0.55\pm 0.16$ & $-105\pm 30$ \\
 4     & $6.6\pm 0.2$ &  $0.30\pm 0.05$ & $146\pm 11$ \\
\hline
\end{tabular}
}
\label{tb:modes16}
\end{table}

It is interesting to stress that the shorter wavelengths ($\sim 35,\,16.5,\,6.6\,{\rm mas}$) listed in Table~\ref{tb:modes16} are around integer fractions of the longest excited mode ($\sim 1/3,\, 1/6$ and $1/15$ of $\sim 100\,{\rm mas}$). This indicates that they could be triggered in the jet as harmonics of the fundamental frequency and suggests that a single triggering mechanism (e.g., jet base precession) could be responsible for the whole set of observed oscillation patterns.

\begin{figure}[t!]
\includegraphics[width=0.48\textwidth]{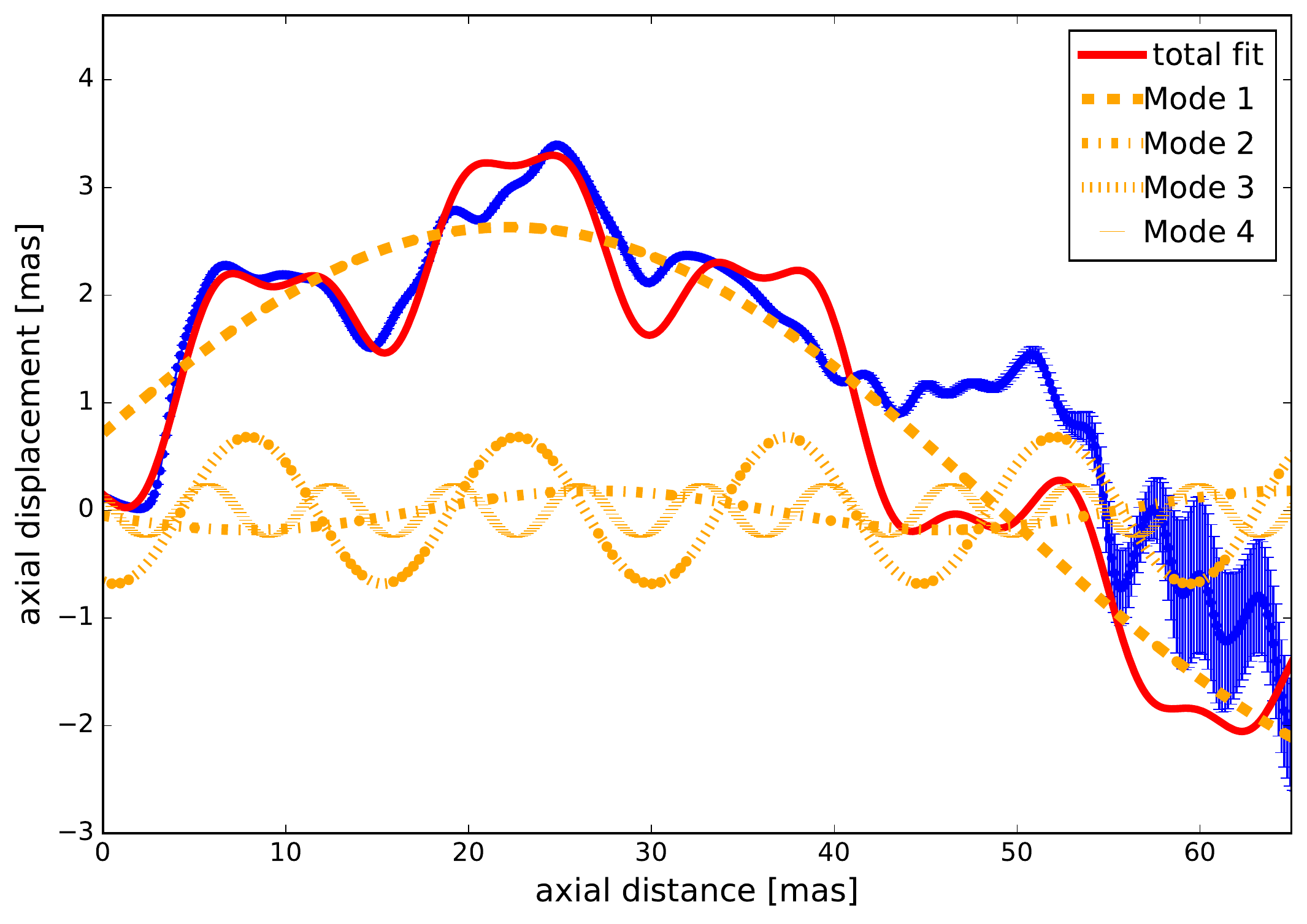}%\,\includegraphics[width=0.48\textwidth]{Modelling/RA16GHz.png}
\caption{Fits by oscillatory modes to the jet ridgeline image of one of the Global VLBI images of S5\,0836+710 at 1.6~GHz observed on 24th October 2013. The red line represents the total, multi-mode fit and the blue lines represent contributions from the individual oscillatory modes as described in the legend. Reproduced from \citealt{VegaGarcia}.
 \label{fig:modesfit}}
\end{figure}

\section{Stability analysis}
\label{sc:method}

The simplest way to explain the observed combination of the oscillations is to assume that all the observed modes grow around the frequencies that correspond to the maximum growth rates for each mode (minimum growth lengths). An initial derivation of the jet parameters can be done using the approximation to the solution of the stability problem given by \cite{hardee2000}. This approximation assumes that the jet is separated from the ambient medium by a contact discontinuity and that the jet Mach number is $>> 1$. The parameters derived using this approach also depend on an \emph{a priori} identification of the modes and on the wave speed taken of the mode propagation. The basic jet parameters are then derived from the following equations. The jet Mach number,

\begin{equation} \label{eq:mach}
    M_\mathrm{j} = \frac{\lambda^*(1-\beta_w\cos{\theta_\mathrm{j}})}{8R_\mathrm{j}\gamma_\mathrm{j}(1-\beta_w/\beta_\mathrm{j})\sin{\theta_\mathrm{j}}}\,,
\end{equation}
where $\lambda^*=\lambda_i(n_i+2m_{b_i}+\frac{1}{2})$ is the characteristic wavelength \citep{lobanov2001}, $\lambda_i$ are the observed (projected) wavelengths and $n_i$ is the azimuthal wavenumber, and $m_{b_i}$ indicates the order of the mode (0 for the surface mode and $m_{b_i}$ for the $m_{b_i}$-th body mode). The jet-to-ambient medium density ratio, 
\begin{equation}
    \eta =\frac{M_\mathrm{j}^2}{M_\mathrm{x}^2},    
\end{equation}
where the external jet Mach number, $M_\mathrm{x}$, and the intrinsic jet pattern speeds are calculated with:
\begin{equation}
    M_\mathrm{x} = \frac{\lambda^*\beta_\mathrm{j}(1-\beta_w\cos{\theta_\mathrm{j}})}{8R_\mathrm{j}\beta_w\sin{\theta_\mathrm{j}}}\,,
\end{equation}
\begin{equation}
\beta_w = \frac{\beta_{w,\mathrm{app}}}{\sin{\theta_\mathrm{j}}+\beta_{w,\mathrm{app}}\cos{\theta_\mathrm{j}}}\,, 
\end{equation}
and
\begin{equation} \label{eq:beta}
    \beta_\mathrm{j} = \frac{\beta_\mathrm{app}}{\sin{\theta_\mathrm{j}}+\beta_\mathrm{app}\cos{\theta_\mathrm{j}}},
\end{equation}
where $R_j$ is the jet radius at the jet base, $\theta_j$ is the viewing angle $\beta_\mathrm{app}$ is the apparent speed and $\beta_{w,\mathrm{app}}$ is the apparent pattern speed. 

In these expressions, $\beta_w$ is also difficult to determine, and it has to be taken into account that each developing mode may have a different wave speed. With all these caveats, using the values given in \cite{otterbein+1998}, i.e., $\theta_\mathrm{j}=3^\circ$ and a bulk Lorentz factor $\gamma_\mathrm{j} = 12$, \cite{lobanov+1998} derived a jet classical Mach number of $M_j = 6$ and a jet-to-ambient medium density ratio of $\eta = 0.04$. The three longest wavelengths have been indentified as corresponding to the surface, first body and second body modes, respectively \citep{VegaGarcia}. This identification results in $M_\mathrm{j} = 12 \pm 3$ and $\eta = 0.33 \pm 0.08$. Differences may be attributed to the different aproaches or to the difficulties in the measure of $\beta_w$. 

In order to overcome these problems and provide a self-consistent framework for jet modelling, we propose an independent method to derive estimates of the jet parameters from the direct solutions of the stability equation. Taking into account that the presence of a shear layer surrounding the jet in S5\,0836+710 has been suggested in earlier works \citep{perucho+2007b,perucho+2011}, we solve the differential equation of the linearized stability problem of two flows in pressure equilibrium and relative velocity, for relativistic, sheared jets in cylindrical coordinates \citep[see, e.g.,][]{Birkinshaw1984,Birkinshaw1991}: 

\begin{equation}
\begin{matrix}
0 \,=\,\dfrac{d^2 p_1(r)}{dr^2}\, +\, \\ \dfrac{d p_1(r)}{dr} \llave{ \dfrac{1}{r} + \dfrac{2 \gamma_{0,\mathrm{j}}(r) ^2 \dfrac{dv_{0,z}(r)}{dr} \left(k - \dfrac{\omega v_{0,z}(r)}{c^2}\right)}{w-k v_{0,z}(r)} - \dfrac{\dfrac{d\rho_0(r)}{dr}}{\rho_0 + \dfrac{p_0}{c^2}}} + \\
p_1(r) \llave{\gamma_{0,\mathrm{j}}(r)^2 \corch{\dfrac{\rho_0(r) (\omega - k v_{0,z}(r))^2}{\Gamma p_0} - \paren{k - \dfrac{\omega v_{0,z}(r)}{c^2}}^2} - \dfrac{n^2}{r^2}},
\end{matrix}
%\nonumber
\label{eq:dif}
\end{equation}
where $r$ and $z$ are the radial and axial coordinates, $k$ and $n$ are the axial and azimuthal components of the wavenumber, respectively, $\omega$ is the frequency, $\Gamma$ is the adiabatic index, $p$ is the pressure, and subscripts $0$ and $1$ refer to unperturbed and perturbed variables, respectively. The perturbations are assumed to be proportional to $ g(r) \exp{(i(k_z z - \omega t))}$, where $g(r)$ describes the radial structure of the wave. Like in previous works using the same approach \citep{perucho+2007a,perucho+2007b}, the shear layer that we consider is described by the variable $a(r)$, corresponding to the velocity or rest-mass density:

\begin{equation}
    a(r) = a_\infty+(a_0-a_\infty)/\cosh{r^m}\,,
    \label{eq:shearlayer}
\end{equation}
where $a_0$ is the value of the parameter at $r=0$, $a_\infty$ its value when $r \rightarrow \infty$, and $m$ defines the steepness of the layer. This has been shown to converge to the vortex-sheet solution for large enough values of $m$ \citep{perucho+2005}.

 We assume an ideal gas equation of state with adiabatic index $\Gamma = 4/3$ to describe both the jet and ambient medium. We take a relativistic hydrodynamics approach, i.e., we implicitly assume that the magnetic field is dynamically negligible at the studied scales. 
Finally, we impose the following boundary conditions: 
 
 \begin{enumerate}[i]
    \item non incoming waves from infinity (Sommerfeld condition), and
    \item symmetry or antisymmetry of the perturbation and its first derivative on the jet axis.
 \end{enumerate}
 
\begin{figure*}[t!]
\includegraphics[width=0.48\textwidth]{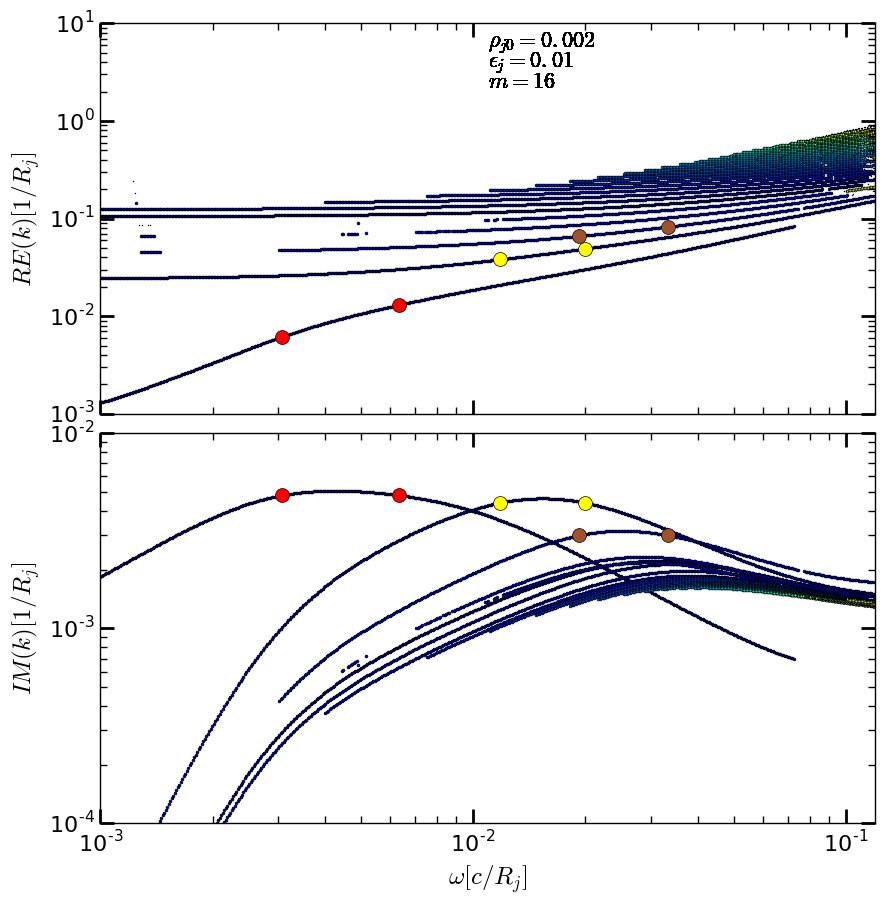}
\includegraphics[width=0.48\textwidth]{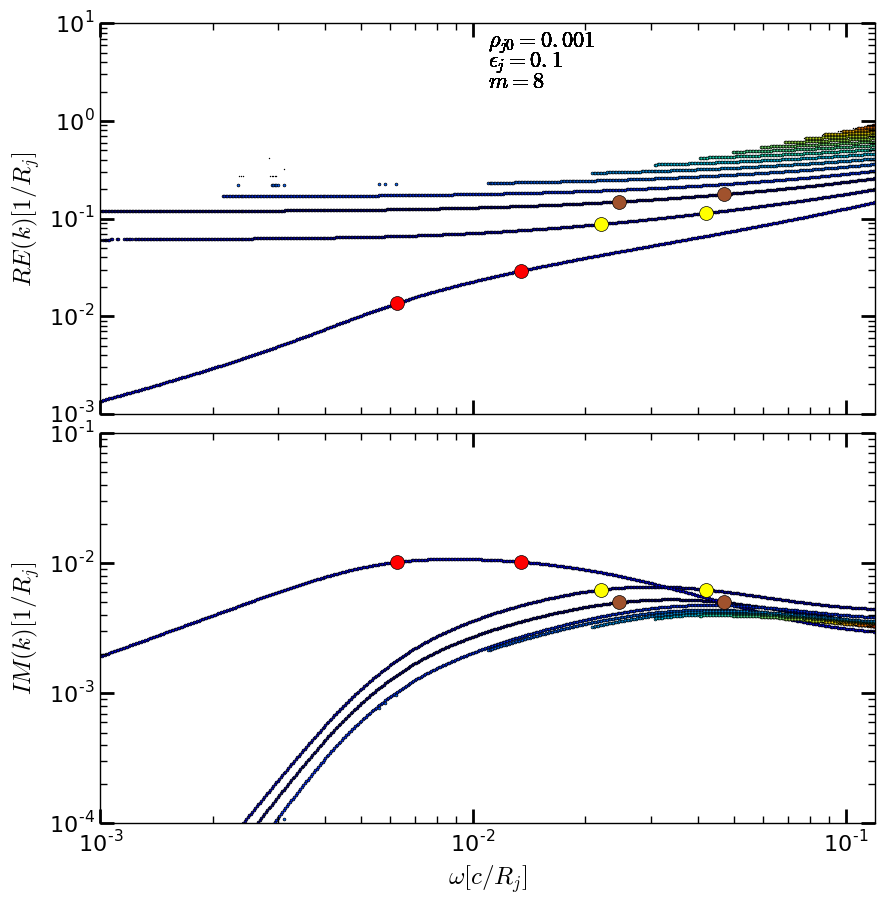}
\caption{Example solutions of the stability equation for helical modes applied to the jet of S5\,0836+710. The upper panel represents the wave-number and the lower panel the growth length as function of the frequency. The coloured dots represent the values for which the maximum value of the imaginary part of the wave-number is reduced by $5\%$, i.e., the interval where the growth length is minimized. The red dots correspond to the helical surface mode, the yellow dots to the first body mode and the brown ones to the second body mode. {\bf Left panel:} Solutions for a shear layer of $m=16$, jet internal energy of $0.01$ and jet density of $0.002$. {\bf Right panel:} Solutions for a shear layer of $m=8$, jet internal energy of $0.01$ and jet density of $0.001$.}
 \label{fig:sim1}
\end{figure*}

Then, we obtain the solutions to the differential equation using the shooting-method \citep[see][for details]{perucho+2005,perucho+2007a}, which provides us with values for $k$ and $\omega$ that satisfy Eq.~\ref{eq:dif}. We used the spatial approach to do this, i.e., we assume that the unstable modes grow in distance, because we study how the perturbations grow downstream along the jet. This approach thus uses real frequencies and complex wave numbers. The real part of the wave number describes the wavelength of the perturbation, $\lambda_{\textrm{int}}$, and the imaginary part describes the growth length, $\lambda_\mathrm{e}$:
 
 \begin{equation}
 \label{eq:waveAndGrow}
   \lambda_{\textrm{int}} = \frac{2\pi}{\Re(k)}\,,
 \end{equation}
 
  \begin{equation}
 \label{eq:waveAndGrow2}
   \lambda_{\mathrm{e}} = - \frac{1}{\Im(k)}.
 \end{equation}
We have solved the equation using a range of shear layer widths, and we have used different values of the Lorentz factor and have swept several orders of magnitude in jet-to-ambient medium density ratios and specific internal energy for each shearing width (see Appendix C for the complete set of solutions obtained). In the calculations, we use units of $\rho_a=1$, $c=1$ and $R_j=1$ (jet radius). Therefore, the jet density in the equations, $\rho$, represents the density ratio, $\eta$, between the jet and the ambient medium. 

Specifically, in our parameter space, $\eta$ ranges from $10^{-5}$ to $10^{-1}$, the jet specific internal energies, $\varepsilon_\mathrm{j}$ take values from 0.001~$c^2$ to $c^2$, i.e., the calculations include both cold and hot jets, the shear layer widths range from a narrow layer, as given by $m=16$ or $m=12$ ($\leq 10$ \% of the jet radius), to broader ones with $m=8$ and $m=4$ ($\simeq 20$ \% of the jet radius). Finally, we considered Lorentz factors from from 5 to 17, including $\gamma_\mathrm{j}$= 12 as derived from VLBI observations at 8 GHz \citep{otterbein+1998}, $\gamma_\mathrm{j}=17$ from VLBI observations at 15 GHz \citep{lister+2013}, and $\gamma_\mathrm{j}=5$ to study the option of a slower layer of plasma surrounding a fast inner spine.

 As a result of the calculations, we obtain a set of physical parameters that provide the best fit to the observed wavelengths, and we can use them to derive an approximation of the Mach number of the flow, using the classical Mach number definition, $M=v_\mathrm{j}/c_\mathrm{sj}$, where the jet sound speed, $c_\mathrm{sj}$, can be calculated as:

\begin{equation}
    c_\mathrm{sj}^2 = \frac{\Gamma p}{\rho h} = \frac{\Gamma (\Gamma -1) \varepsilon_\mathrm{j}}{1 + \Gamma \varepsilon_\mathrm{j}}\,,
\end{equation}
where $h= (1 + \Gamma \varepsilon$) is the specific enthalpy.

In the lower panels of Fig.~\ref{fig:sim1}, we indicate the maxima of the imaginary part of the wavenumber (or minimum growth lengths) for each mode. This allows us to derive the intrinsic wavelength of the modes by looking for the maximum in values of the wavenumber in the upper plots. Since the curves do not have sharp maxima, but plateau-like maxima, we consider the points where the maximum value of the imaginary part for each mode of the wave-number is reduced by $5\%$ as lower and upper bounds. In order to compare our results with the fitted structures from observations, we convert the distance unit used in the calculations (jet radius, $R_j$) into milliarcseconds, using the jet radius at the jet base derived from observations ($1\,R_j = 2\,{\rm mas}$, \citealt{VegaGarcia}). In this respect, we note that the stability problem is solved for an infinite, cylindrical jet, but jet expansion introduces an increase of the mode peak wavelengths \citep[e.g.,][]{Hardee1982,Hardee1984,Hardee1986} so our assumption represents a source of uncertainty. However, the jet opening angle for the jet in  S5\,0836+710 is very small \citep[$< 1^\circ$, see][]{perucho+2012a}, and we can regard our solutions as a plausible approach to the jet parameters.
 
The corresponding observed wavelength can be then derived from the calculated intrinsic
wavelength, using the viewing angle, $\theta$, and the intrinsic wave speed, $\beta_\omega$:

\begin{equation}
    \label{eq:obslambda}
    \lambda_{\textrm{obs}} = \lambda_{\textrm{int}} \frac{\sin{\theta}}{1-\beta_\omega \cos{\theta}}.
\end{equation}
The intrinsic wave speed has been estimated using the values of the complex $k$ and real $\omega$ from the calculations using:

\begin{equation}
    \label{eq:vintrinsic}
    \beta_{\omega} = \frac{\omega}{k} \sim  \Re\left(\frac{\omega}{k}\right),
\end{equation}
using that $\Re(k) \gg \Im (k)$.

\begin{figure*}[t!]
\centering
	\includegraphics[width=0.48\textwidth]{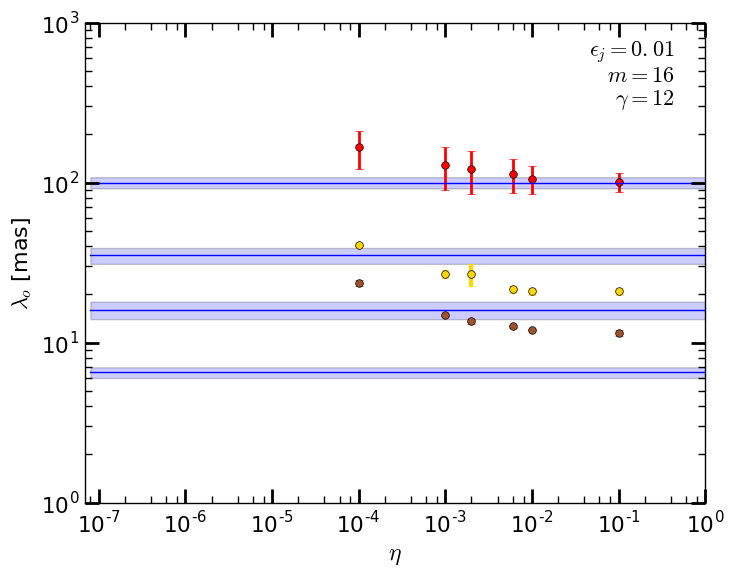}
	\includegraphics[width=0.48\textwidth]{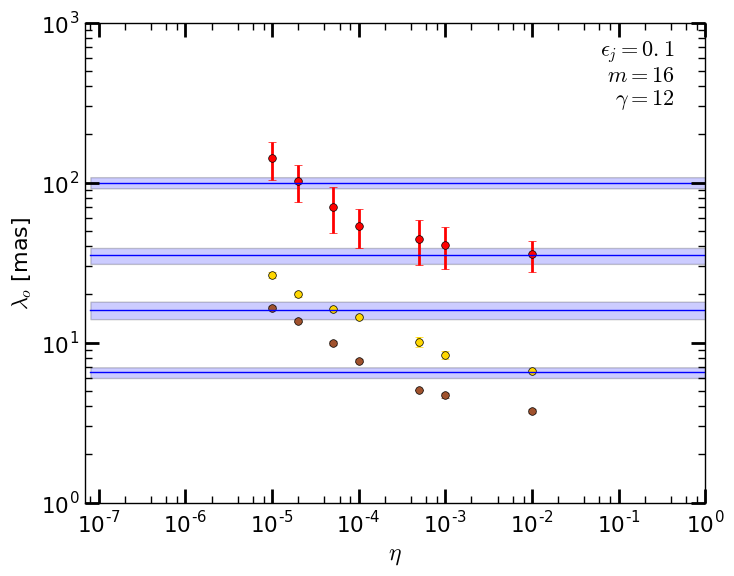}
	\includegraphics[width=0.48\textwidth]{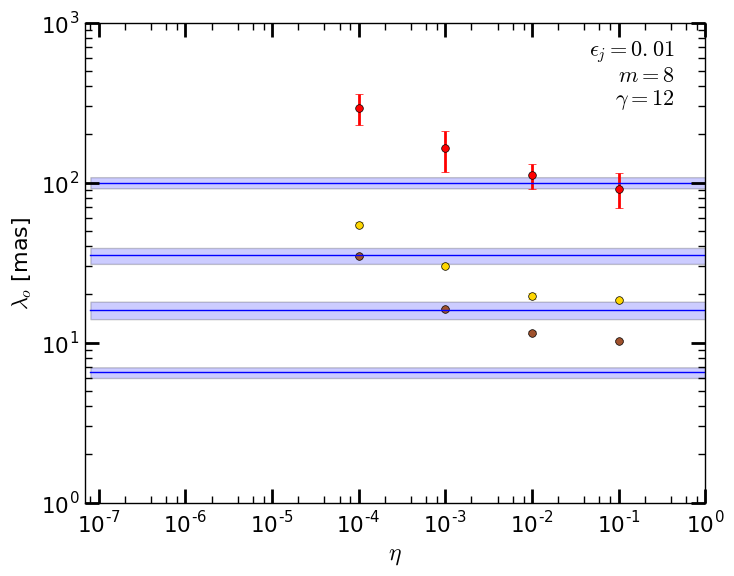}
	\includegraphics[width=0.48\textwidth]{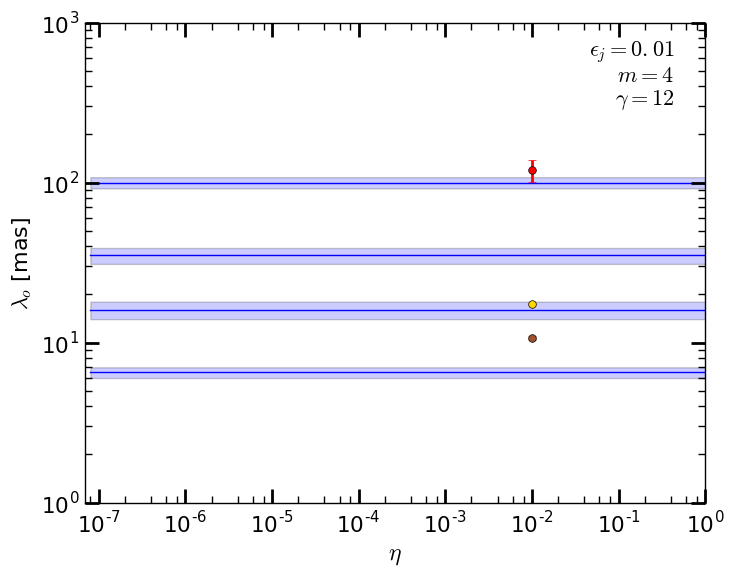}
	\caption{Observed wavelengths  and their errors (blue lines and shades) compared to the wavelengths of different instability modes (symbols) obtained from the linear stability analysis. The color scheme used for representing the calculated wavelengths is the same as the one used in Fig~\ref{fig:sim1}. The $x-$axis represents the density ratio and the $y-$axis the corresponding observed wavelength. {\bf Upper left panel:} Results for a shear layer of steepness $m=16$ and jet internal energy of $0.01$.  {\bf Upper right panel:} Results for a shear layer of steepness $m=16$ and jet internal energy of $0.1$. {\bf Lower left panel:} Results for a shear layer of steepness $m=8$ and jet internal energy of $0.01$.  {\bf Lower right panel:} Results for a shear layer of steepness $m=4$ and jet internal energy of $0.01$.}
	\label{fig:simresults}
\end{figure*}

\section{Results: Jet parameters}
\label{sc:jetparameters}

The values of the peak wavelengths that result from the solution of Eq.~\ref{eq:dif} for each set of parameters and the Mach numbers derived from them can be found in the tables in Appendix~\ref{ap:simulations}. In order to illustrate how the predicted observed wavelengths change with the shear layer width and the jet parameters, we have plotted in Fig.~\ref{fig:simresults} the wavelengths obtained for the relevant modes as a function of rest-mass density ratio for each value of the shear layer and jet internal energy. This figure only shows the best result (given by the most probable set of jet parameters) and other three interesting examples. Additional figures can be found in the Appendix~\ref{ap:simulations}. In the plots, the horizontal lines and shades indicate the observed wavelengths and their formal errors. Each \emph{family} of three points in the $y$ direction represents the wavelengths at the maxima in growth rate of the fundamental, first and second body modes for each different model (defined by the given set of physical parameters). The most likely model is thus the one for which the points lie closer to the observed wavelengths (horizontal lines) within the error given by the shaded areas.

If we first compare the effect of the shear layer on the simulated observed wavelengths, we can see that the broader the shear layer, the larger the separation between the peaks of the fundamental mode and the first body mode, while the narrower the shear layer, the smaller the separation between these modes. This ratio excludes wide shear-layers extending more than $1\,R_j$ (e.g., $m \geq 4$). The ratio between the observed wavelengths indicates a narrower shear layer width $m=16$ (with an extension of $0.1-0.2\,R_j$) as a most likely scenario. Regarding the effect of the physical parameters of the jet, an increase of the density ratio reduces the peak wavelengths, since it causes an increase of the specific internal energy, whereas increasing the Lorentz factor increases those wavelengths. These solid trends have allowed us to exclude wide regions of the parameter space.

The dispersion between the observed wavelengths and the calculated ones can be used to determine which of the parameter sets best recovers the observed values given in Section~\ref{sc:ridgeLineOscillations}. The two models with lower dispersions are the ones with $\eta = 2 \times 10^{-3}$ and $\varepsilon_\mathrm{j} = 0.01c^2$, and $\eta = 10^{-3}$ and $\varepsilon_\mathrm{j} = 0.01c^2$. Both cases would correspond to a Mach number of $M \sim 12$. This value coincides with the value obtained independently from the approximations used in previous works (see Section~\ref{sc:method}). The specific internal energy of the jet is clearly lower than one ($c^2$), which places the jet in the cold regime, possibly dominated by the kinetic energy. 

We have also computed the solutions for different jet Lorentz factors values, $\gamma_\mathrm{j} = 5$ and $\gamma_\mathrm{j} = 17$, for the best fitting values of $\varepsilon_\mathrm{j}$ and $\eta$. We found that the case of $\gamma_\mathrm{j} = 17$ results in peak wavelengths that are compatible with the observed wavelengths for each of the helical modes proposed (surface, and first and second body modes), whereas the $\gamma_\mathrm{j} = 5$ models systematically result in lower peak wavelengths for each of the modes. Comparing the results given by Lorentz factors 12 and 17, we see that although both models could still be compatible, the dispersion between the calculated wavelengths and the observed ones is lower for the Lorentz factor of 12 (see the Tables in the Appendix). It is noteworthy that the parameter sets that give the lowest dispersion for Lorentz factor 17 include options with larger values of $\varepsilon_\mathrm{j}$ ($0.01 - 0.1\,c^2$) and smaller $\eta$ ($10^{-5}-10^{-4}$). Finally, although we found a relatively small dispersion values for paremeter sets with $\gamma =12$ and higher values of $\varepsilon_\mathrm{j}$ ($1 - 10\,c^2$), the density ratios are extremely small ($10^{-7}-10^{-6}$), and still the dispersion is larger than for the aforementioned parameter sets. 

Our results thus allow us to confirm that the Lorentz factor of the jet at the scales revealed by the $1.6\,$GHz jet is probably in the range $12-17$, i.e., in agreement with the estimated value by \citet{otterbein+1998} from jet kinematics at 8~GHz and that derived by \cite{lister+2013} from 15~GHz observations. For this values of the Lorentz factor, we find that the jet is remarkably underdense with respect to its environment ($\eta \sim 10^{-5}-10^{-3}$) and relatively cold ($\varepsilon_\mathrm{j} \sim 0.01 - 0.1\,c^2$). This results in a relativistic jet Mach number, $\mathcal{M}_\mathrm{j} \simeq \gamma_\mathrm{j}M_\mathrm{j}$, range of $60 - 144$. 

%The difference with the Lorentz factor given by 15~GHz observations \citep{lister+2013} may be explained by this frequency revealing inner jet regions, and to transversal jet structure.

%We can use the solutions of the stability problem to perform a sanity check of the characteristic wavelength relation discussed in Section~\ref{sc:method}. In order to do this, we introduce the obtained parameter sets in the corresponding equations (Eqs.~\ref{eq:mach}-\ref{eq:beta}) and compute the corresponding characteristic wavelengths, $\lambda^*$. We find similar values of $\lambda^*$ for the body modes ($n=1,\,m>0$) to those given by the numerical solutions. In contrast, we obtain much longer characteristic wavelengths for the fundamental mode than obtained from the stability equation (Eq.~\ref{eq:dif}). Nevertheless, once the first body mode has been identified, any obviously longer wavelength corresponds necessarily to the fundamental mode, so this identification can also be taken as accurate. Interestingly, both independent approaches result in similar Mach numbers of the jet, albeit with very different density ratios. 

\section{Discussion and conclusions}
\label{sc:summary}

Our calculations include the following basic assumptions: 1) infinite cylindrical jet, and 2) dynamically negligible magnetic field. Because we model the observed jet structures over $> 100\,{\rm mas}$ in length, representing linear scales of $\sim 10\,{\rm kpc}$, and recalling that the jet flow may change considerably along this distance, we have to discuss the validity of our results as estimates of mean jet parameters. On the one hand, taking into account that the analysis pictures a cold and fast jet, and that our estimate of the Lorentz factor coincides with that derived by \cite{otterbein+1998} from jet kinematics at 8~GHz, we can infer that any kind of Bernoulli acceleration (internal energy conversion into kinetic energy) must have taken place upstream and that no further acceleration takes place along the observed jet at 1.6 GHz. This argumentation consolidates the value of the Lorentz factor of $\gamma \simeq 12$ for the whole VLBI jet in S5\,0836+710. The high degree of collimation of this jet also allows us to take the derived values of the specific internal energy and density ratio can be regarded as an order of magnitude estimate because the limited jet expansion will lead to a relatively small drop in both parameters. As far as the dynamical role of the magnetic field, magnetic acceleration models \citep[e.g.,][]{vlahakis+2004,komissarov2012} and observational \citep[e.g.,][]{homan+2015,boccardi+2016,mertens2016} results seem to point towards magnetic acceleration taking place at the inner tens of parsecs, at most, from injection, so a magnetically dominated jet is not expected at kiloparsec scales, as those involved in the VLBI jet of S5\,0836+710. In addition, our results tend to favor a kinetically dominated jet, with relatively low internal energy and high Lorentz factor \citep[as expected for particle dominated flows,][]{marti+2016}. This can be taken as a self-consistency test for our initial hypothesis of the observed structures being generated by KH instability modes. The similarity between the inferred Lorentz factor and that given by previous kinematic studies \citep{otterbein+1998,lister+2013} validates our approach and encourages us to propose it as a possible way to derive jet parameters in other cases, for the range of distances in which the assumptions are fulfilled. Namely, small jet opening angles and linear-growth regime of the modes (implying relatively small changes in the wave amplitude). 

Because the calculations assume that the jet parameters are constant in the axial direction, we have to consider the parameters derived as mean values along the observed VLBI jet. Further work and should include genetic/machine learning algorithms in the application of this method to the determination of jet parameters. 

The values derived for the jet Lorentz factor and Mach number result in a relativistic Mach number of $ \mathcal{M}_\mathrm{j} \sim 60-144$ (c.f. the classical Mach number $M_\mathrm{j} = 5-12$). The jet would thus be well within the stable region of the relativistic jet stability plane shown in \cite[][Fig. 21 in that paper]{perucho+2005}, with very small growth-rate unstable modes (which validates the modelling assumption of constant wave amplitudes in Eq.~\ref{eq:deltaz}) and prone to the development of stabilising, short-wavelength resonant modes \citep{perucho+2007a,Perucho2010}. This is probably in contradiction with the conclusion of \cite{perucho+2012b}, where the authors claim that the jet may be disrupted by the growth of instabilities, giving rise to the decollimated structure observed with MERLIN at arcsecond scales. This structure was interpreted by the authors as a sign of jet disruption, but this work and recent observational results (Kappes et al., in preparation) may indicate that the arcsecond emission region related to this source may be revealing the interaction site of the jet with the intracluster medium.

The results obtained in this work can be summarized as follows:
\begin{itemize}
\item We have used the results from previous studies of the S5\,0836+710 jet structure that have modelled its ridgeline at 1.6\,GHz, which covers up to $\simeq 130\,{\rm mas}$, by fitting transverse brightness profiles with a single Gaussian component. 
\item The ridgeline shows the presence of several oscillatory terms, which have been reported in several different works \citep{lobanov+1998,lobanov+2006,perucho+2012a}. All those previous works give similar values for the wavelengths of the most prominent oscillations, using different arrays and epochs of observation. 
\item We assume that these oscillations correspond to KH instability modes developing in the flow, and propose a method to derive the physical parameters of the jet. The method requires solving numerically the stability problem for a wide range of values of the relevant jet parameters and comparing the wavelengths at the maximum growth-rates for the identified modes plus their ratios with the observed structures.
\item This method does not need any \emph{a priori} known parameters, but only solving the linear stability problem for a wide region of the parameter space. From this approach, we find a jet Lorentz factor in the range $\gamma\simeq 12-17$, in coincidence with kinematical studies \citep{otterbein+1998,lister+2013}, and, within that range, the jet-to-ambient medium density ratio is found to be $\eta \sim 10^{-5} - 10^{-3}$, and the jet specific internal energy is $\varepsilon \sim 10^{-2}- 10^{-1}\, c^2$, resulting in a jet classical Mach number of $M \sim 5-12$. We also found that the observed radio jet is surrounded by a thin shear layer, with a width $\simeq 10$~\% of the jet radius. The value derived for the specific internal energy and the Lorentz factor point towards a kinetically dominated jet.
\end{itemize}

\section*{Acknowledgments}

L.V.G is a member of the International Max Planck Research
School (IMPRS) for Astronomy and Astrophysics at the Universities of Bonn
and  Cologne. This research is based
on observations correlated at the Bonn Correlator, jointly operated by
the Max Planck Institute for Radio Astronomy (MPIfR), and the Federal
Agency for Cartography and Geodesy (BKG).  The European VLBI Network
is a joint facility of European, Chinese, South African and other
radio astronomy institutes funded by their national research
councils. The National Radio Astronomy Observatory is a facility of
the National Science Foundation operated under cooperative agreement
by Associated Universities, Inc.  Thanks to Phillip Edward, Alan Roy, and V\'ictor M. Pati\~no \'Alvarez for the useful comments about the paper. We thank the anonymous referee for useful comments which improved the manuscript. MP has been supported by the Spanish Ministerio de Econom\'{\i}a y
Competitividad (grants AYA2015-66899-C2-1-P and AYA2016-77237-C3-3-P)
and the Generalitat Valenciana (grant PROMETEOII/2014/069). The authors want to thank P.E. Hardee for his inspiring work.

\bibliographystyle{aa}
%\bibliography{ra0836.bib}

\begin{appendix}

 \section{Results of the calculations for the stability studies}
 \label{ap:simulations}
 
 Here we present the table with all the values of the simulated observed wavelengths and the estimated Mach number values for all the calculations presented throughout the paper. We also present the remaining plots of $\lambda_\mathrm{obs}$ as a function of the density ratio obtained from this analysis (Section~\ref{sc:jetparameters}).

\begin{table*}[h!]
\caption{Solutions of the dispersion relation for $m=16$ and $\gamma=12$}
\begin{tabular}{c|cccccccc}%
Mode&m&$\varepsilon_\mathrm{j}$&$\eta$&$M_\mathrm{j}$&$\omega$ [$c/R_\mathrm{j}$]&$\Re(k)$ [$1/R_\mathrm{j}$]&
$\lambda$ [mas] & Dispersion\\%
\hline%
\hline%

$H_{s}$&16&10&$10^{-7}$&1.55&0.0019$-$0.0052&0.0054$-$0.0117&96.69$-$167.61 & \\ %
$H_{b_1}$&16&10&$10^{-7}$&1.55&0.0053$-$0.014&0.0310$-$0.0380&23.82$-$25.59 & 0.135\\ %
$H_{b_2}$&16&10&$10^{-7}$&1.55&0.0070$-$0.0120&0.0520$-$0.0590&14.00$-$14.62 & \\%

\hline%
\hline%
$H_{s}$&16&1&$10^{-6}$&1.94&0.0018$-$0.0050&0.0052$-$0.0110&103.97$-$171.92 & \\%
$H_{b_1}$&16&1&$10^{-6}$&1.94&0.0050$-$0.010&0.0310$-$0.0380&23.48$-$25.29 & 0.171\\%
$H_{b_2}$&16&1&$10^{-6}$&1.94&0.0060$-$0.0125&0.0500$-$0.0590&14.15$-$14.96 & \\%
\hline

$H_{s}$&16&1&$10^{-5}$&1.94&0.0055$-$0.0150&0.0160$-$0.0340&32.91$-$55.65 & \\%
$H_{b_1}$&16&1&$10^{-5}$&1.94&0.0150$-$0.0280&0.0900$-$0.1100&8.02$-$8.77 & 0.774\\%
$H_{b_2}$&16&1&$10^{-5}$&1.94&0.0205$-$0.0400&0.1300$-$0.1600&5.48$-$6.01 & \\%
\hline

$H_{s}$&16&1&$10^{-4}$&1.94&0.012$-$0.034&0.033$-$0.076&14.7$-$26.7 & \\%
$H_{b_1}$&16&1&$10^{-4}$&1.94&0.050$-$0.090&0.180$-$0.230&4.69$-$5.05 & 1.207\\%
$H_{b_2}$&16&1&$10^{-4}$&1.94&0.050$-$0.090&0.310$-$0.365&2.39$-$2.53 & \\% 
\hline

$H_{s}$&16&1&0.001&1.94&0.023$-$0.056&0.046$-$0.105&12.23$-$21.57 & \\%
$H_{b_1}$&16&1&0.001&1.94&0.086$-$0.151&0.310$-$0.400&2.63$-$2.93 & 1.358\\%
$H_{b_2}$&16&1&0.001&1.94&0.127$-$0.221&0.520$-$0.630&1.61$-$1.67 & \\%

\hline%
\hline%

$H_{s}$&16&0.1&$10^{-5}$&4.21&0.0018$-$0.0050&0.0050$-$0.0110&103.97$-$180.10 & \\%
$H_{b_1}$&16&0.1&$10^{-5}$&4.21&0.0050$-$0.010&0.0290$-$0.0360&25.28$-$27.40 & 0.187\\%
$H_{b_2}$&16&0.1&$10^{-5}$&4.21&0.0060$-$0.0110&0.0450$-$0.0520&16.05$-$16.87 & \\%
\hline

$H_{s}$&16&0.1&$2\times 10^{-5}$&4.21&0.0025$-$0.0067&0.0070$-$0.0150&75.20$-$128.62 & \\%
$H_{b_1}$&16&0.1&$2\times 10^{-5}$&4.21&0.0064$-$0.0120&0.0380$-$0.0460&19.34$-$20.81 & 0.183\\%
$H_{b_2}$&16&0.1&$2\times 10^{-5}$&4.21&0.010$-$0.0220&0.0570$-$0.0710&13.42$-$14.00 & \\%
\hline

$H_{s}$&16&0.1&$5\times 10^{-5}$&4.21&0.0040$-$0.010&0.010$-$0.0230&48.22$-$93.20 & \\%
$H_{b_1}$&16&0.1&$5\times 10^{-5}$&4.21&0.010$-$0.0190&0.0500$-$0.0600&16.04$-$16.45 & 0.332\\%
$H_{b_2}$&16&0.1&$5\times 10^{-5}$&4.21&0.0130$-$0.0260&0.0770$-$0.0940&9.67$-$10.28 & \\%
\hline

$H_{s}$&16&0.1&$10^{-4}$&4.21&0.0045$-$0.0126&0.0129$-$0.0286&38.88$-$68.18 & \\%
$H_{b_1}$&16&0.1&$10^{-4}$&4.21&0.0160$-$0.0330&0.0600$-$0.0800&13.98$-$14.94 & 0.487\\%
$H_{b_2}$&16&0.1&$10^{-4}$&4.21&0.0170$-$0.0325&0.1$-$0.1200&7.51$-$7.92 & \\%
\hline

$H_{s}$&16&0.1&$5\times 10^{-4}$&4.21&0.0060$-$0.0160&0.0150$-$0.0360&30.45$-$58.02 & \\%
$H_{b_1}$&16&0.1&$5\times 10^{-4}$&4.21&0.0300$-$0.0500&0.0900$-$0.1200&9.35$-$10.90 & 0.755\\%
$H_{b_2}$&16&0.1&$5\times 10^{-4}$&4.21&0.0300$-$0.0600&0.1600$-$0.1900&5.06$-$5.06 & \\%
\hline

$H_{s}$&16&0.1&0.001&4.21&0.007$-$0.017&0.017$-$0.038&28.71$-$53.06 & \\%
$H_{b_1}$&16&0.1&0.001&4.21&0.033$-$0.057&0.107$-$0.140&7.88$-$8.84 & 0.972\\%
$H_{b_2}$&16&0.1&0.001&4.21&0.050$-$0.090&0.186$-$0.236&4.50$-$4.84 & \\%
\hline

$H_{s}$&16&0.1&0.01&4.21&0.019$-$0.035&0.030$-$0.056&27.79$-$43.23 & \\%
$H_{b_1}$&16&0.1&0.01&4.21&0.063$-$0.110&0.160$-$0.210&6.56$-$6.77 & 0.972\\%
$H_{b_2}$&16&0.1&0.01&4.21&0.080$-$0.140&0.250$-$0.320&3.65$-$3.87 & \\%

\hline%
\hline%

$H_{s}$&16&0.01&$10^{-4}$&12.53&0.0015$-$0.0041&0.0042$-$0.0092&121.71$-$211.18 & \\%
$H_{b_1}$&16&0.01&$10^{-4}$&12.53&0.0050$-$0.010&0.0200$-$0.0260&39.00$-$41.94 & 0.499\\%
$H_{b_2}$&16&0.01&$10^{-4}$&12.53&0.0054$-$0.010&0.0320$-$0.0390&22.68$-$24.73 & \\%
\hline

$H_{s}$&16&0.01&$10^{-3}$&12.53&0.0024$-$0.0055&0.0054$-$0.0122&89.69$-$166.25 & \\%
$H_{b_1}$&16&0.01&$10^{-3}$&12.53&0.0110$-$0.0190&0.0340$-$0.0450&25.14$-$28.41 & 0.096\\%
$H_{b_2}$&16&0.01&$10^{-3}$&12.53&0.0160$-$0.0310&0.0600$-$0.0760&14.61$-$14.95 & \\%
\hline

$H_{s}$&16&0.01&$2\times 10^{-3}$&12.53&0.0030$-$0.0060&0.0060$-$0.0130&85.19$-$157.74 & \\%
$H_{b_1}$&16&0.01&$2\times 10^{-3}$&12.53&0.0120$-$0.0200&0.0330$-$0.0490&22.55$-$31.01 & 0.075\\%
$H_{b_2}$&16&0.01&$2\times 10^{-3}$&12.53&0.0190$-$0.0330&0.0660$-$0.0830&13.14$-$13.99 & \\%
\hline

$H_{s}$&16&0.01&$6\times 10^{-3}$&12.53&0.0045$-$0.0084&0.0078$-$0.0152&85.60$-$140.39 & \\%
$H_{b_1}$&16&0.01&$6\times 10^{-3}$&12.53&0.0150$-$0.0250&0.0440$-$0.0570&20.47$-$22.59 & 0.155\\%
$H_{b_2}$&16&0.01&$6\times 10^{-3}$&12.53&0.0210$-$0.0360&0.0720$-$0.0890&12.40$-$12.89 & \\%
\hline

$H_{s}$&16&0.01&0.01&12.53&0.006$-$0.011&0.01$-$0.018&84.37$-$126.43 & \\%
$H_{b_1}$&16&0.01&0.01&12.53&0.019$-$0.033&0.049$-$0.066&19.86$-$21.85 & 0.174 \\%
$H_{b_2}$&16&0.01&0.01&12.53&0.025$-$0.045&0.079$-$0.100&11.95$-$12.18 & \\%
\hline

$H_{s}$&16&0.01&0.1&12.53&0.015$-$0.022&0.020$-$0.029&87.23$-$114.63 & \\%
$H_{b_1}$&16&0.01&0.1&12.53&0.047$-$0.078&0.077$-$0.111&19.85$-$21.84 & 0.183\\%
$H_{b_2}$&16&0.01&0.1&12.53&0.070$-$0.120&0.125$-$0.180&10.94$-$11.94 & \\%

\hline%
\hline%

$H_{s}$&16&0.001&0.1&39.36&0.005$-$0.007&0.006$-$0.009&302.69$-$520.51 & \\%
$H_{b_1}$&16&0.001&0.1&39.36&0.016$-$0.021&0.025$-$0.031&65.50$-$72.76 & 9.963\\%
$H_{b_2}$&16&0.001&0.1&39.36&0.027$-$0.035&0.045$-$0.051&41.00$-$36.48 & \\%

\end{tabular}%
\label{tab:m16}
\end{table*}

\begin{table*}[h!]
\caption{Solutions of the dispersion relation for $m=8$ and $\gamma=12$}
\begin{tabular}{c|cccccccc}%
Mode&m&$\varepsilon_\mathrm{j}$&$\eta$&$M_\mathrm{j}$&$\omega$ [$c/R_\mathrm{j}$]&$\Re(k)$ [$1/R_\mathrm{j}$]&
$\lambda$ [mas] & Dispersion\\%
\hline%
\hline%
$H_{s}$&8&1&$10^{-6}$&1.94&0.0010$-$0.0020&0.0030$-$0.0052&195.02$-$296.18 & \\%
$H_{b_1}$&8&1&$10^{-6}$&1.94&0.0030$-$0.0044&0.0160$-$0.0190&45.03$-$50.57 & 2.166\\%
$H_{b_2}$&8&1&$10^{-6}$&1.94&0.0035$-$0.0047&0.0280$-$0.0300&26.01$-$26.86 & \\%
\hline

$H_{s}$&8&1&$10^{-5}$&1.94&0.0024$-$0.0071&0.0065$-$0.0164&66.41$-$131.81 & \\%
$H_{b_1}$&8&1&$10^{-5}$&1.94&0.0080$-$0.0160&0.0500$-$0.0620&14.29$-$15.66 & 0.384\\%
$H_{b_2}$&8&1&$10^{-5}$&1.94&0.010$-$0.0200&0.0820$-$0.0970&8.54$-$9.14 & \\%
\hline

$H_{s}$&8&1&$10^{-4}$&1.94&0.007$-$0.019&0.018$-$0.044&24.59$-$48.24 & \\%
$H_{b_1}$&8&1&$10^{-4}$&1.94&0.024$-$0.047&0.126$-$0.156&6.03$-$6.45 & 0.989\\%
$H_{b_2}$&8&1&$10^{-4}$&1.94&0.030$-$0.060&0.205$-$0.249&3.48$-$3.76 & \\%
\hline

$H_{s}$&8&1&0.001&1.94&0.019$-$0.038&0.038$-$0.077&15.33$-$26.25 & \\%
$H_{b_1}$&8&1&0.001&1.94&0.061$-$0.114&0.265$-$0.336&2.96$-$3.22 & 1.307\\%
$H_{b_2}$&8&1&0.001&1.94&0.083$-$0.161&0.445$-$0.540&1.74$-$1.82 & \\% 

\hline%
\hline%
$H_{s}$&8&0.1&$10^{-5}$&4.21&0.0010$-$0.0022&0.0027$-$0.0052&206.16$-$335.70 & \\%
$H_{b_1}$&8&0.1&$10^{-5}$&4.21&0.0026$-$0.0053&0.0160$-$0.0200&44.57$-$49.07 & 2.956\\%
$H_{b_2}$&8&0.1&$10^{-5}$&4.21&0.0032$-$0.0064&0.0270$-$0.0320&25.70$-$27.65 & \\%
\hline

$H_{s}$&8&0.1&$10^{-4}$&4.21&0.0024$-$0.0066&0.0060$-$0.0150&73.34$-$146.59 & \\%
$H_{b_1}$&8&0.1&$10^{-4}$&4.21&0.0075$-$0.0140&0.0440$-$0.0530&16.86$-$18.02 & 0.269\\%
$H_{b_2}$&8&0.1&$10^{-4}$&4.21&0.010$-$0.0210&0.0680$-$0.0820&10.78$-$11.35 & \\%
\hline

$H_{s}$&8&0.1&0.001&4.21&0.006$-$0.013&0.014$-$0.029&37.84$-$65.61 & \\%
$H_{b_1}$&8&0.1&0.001&4.21&0.022$-$0.042&0.090$-$0.110&9.65$-$9.66 & 0.738\\%
$H_{b_2}$&8&0.1&0.001&4.21&0.025$-$0.047&0.150$-$0.180&4.95$-$5.26 & \\%
\hline

$H_{s}$&8&0.1&0.01&4.21&0.016$-$0.027&0.028$-$0.048&28.09$-$42.24 & \\%
$H_{b_1}$&8&0.1&0.01&4.21&0.044$-$0.080&0.150$-$0.190&5.97$-$6.20 & 1.003\\%
$H_{b_2}$&8&0.1&0.01&4.21&0.051$-$0.091&0.230$-$0.280&3.48$-$3.67 & \\%

\hline%
\hline%
$H_{s}$&8&0.01&$10^{-4}$&12.53&0.0010$-$0.0021&0.0026$-$0.0048&227.82$-$355.68 & \\%
$H_{b_1}$&8&0.01&$10^{-4}$&12.53&0.0024$-$0.0045&0.0140$-$0.0170&52.60$-$56.70 & 3.931\\%
$H_{b_2}$&8&0.01&$10^{-4}$&12.53&0.0032$-$0.0068&0.0220$-$0.0260&34.26$-$35.00 & \\%
\hline

$H_{s}$&8&0.01&$10^{-3}$&12.53&0.0020$-$0.0043&0.0044$-$0.0095&116.13$-$210.87 & \\%
$H_{b_1}$&8&0.01&$10^{-3}$&12.53&0.0070$-$0.0140&0.0280$-$0.0364&29.31$-$31.28 & 0.404\\%
$H_{b_2}$&8&0.01&$10^{-3}$&12.53&0.0080$-$0.0150&0.0480$-$0.0565&15.85$-$16.45 & \\%
\hline

$H_{s}$&8&0.01&0.01&12.53&0.005$-$0.009&0.009$-$0.015&90.54$-$130.82 & \\%
$H_{b_1}$&8&0.01&0.01&12.53&0.014$-$0.025&0.046$-$0.060&18.76$-$20.52 & 0.208\\%
$H_{b_2}$&8&0.01&0.01&12.53&0.017$-$0.030&0.073$-$0.088&11.33$-$11.74 & \\%
\hline

$H_{s}$&8&0.01&0.1&12.53&0.012$-$0.017&0.017$-$0.026&69.32$-$114.11 & \\%
$H_{b_1}$&8&0.01&0.1&12.53&0.035$-$0.056&0.070$-$0.092&17.98$-$18.76 & 0.260\\%
$H_{b_2}$&8&0.01&0.1&12.53&0.043$-$0.078&0.106$-$0.143&10.11$-$10.44 & \\% 

\hline%
\hline%

$H_{s}$&8&0.001&0.1&39.36&0.004$-$0.005&0.005$-$0.008&246.47$-$507.97 & \\%
$H_{b_1}$&8&0.001&0.1&39.36&0.012$-$0.016&0.023$-$0.027&59.67$-$59.69 & 7.850\\%
$H_{b_2}$&8&0.001&0.1&39.36&0.023$-$0.030&0.040$-$0.050&32.83$-$38.62 & \\%

\end{tabular}%
\label{tab:m8}
\end{table*}

\begin{table*}[htpb!]
\caption{Solutions of the dispersion relation for $m=4$ and $\gamma=12$}
\begin{tabular}{c|cccccccc}%
Mode&m&$\varepsilon_\mathrm{j}$&$\eta$&$M_\mathrm{j}$&$\omega$ [$c/R_\mathrm{j}$]&$\Re(k)$ [$1/R_\mathrm{j}$]&
$\lambda$ [mas] & Dispersion\\%
\hline%
\hline%
$H_{s}$&4&1&0.0001&1.94&0.005$-$0.01&0.015$-$0.023&47.11$-$59.81 & \\%
$H_{b_1}$&4&1&0.0001&1.94&0.013$-$0.028&0.086$-$0.108&8.22$-$9.01 & 0.765\\%
$H_{b_2}$&4&1&0.0001&1.94&0.016$-$0.035&0.140$-$0.170&4.87$-$5.31 & \\%
\hline
\hline

$H_{s}$&4&0.1&0.001&4.21&0.005$-$0.009&0.01$-$0.020&55.56$-$101.93 & \\%
$H_{b_1}$&4&0.1&0.001&4.21&0.013$-$0.026&0.070$-$0.090&10.27$-$11.54 & 0.606\\%
$H_{b_2}$&4&0.1&0.001&4.21&0.015$-$0.033&0.120$-$0.140&6.15$-$6.27 & \\%
\hline

$H_{s}$&4&0.01&0.01&12.53&0.004$-$0.007&0.008$-$0.013&100.35$-$138.85 & \\%
$H_{b_1}$&4&0.01&0.01&12.53&0.01$-$0.019&0.047$-$0.058&16.86$-$17.78 & 0.280\\%
$H_{b_2}$&4&0.01&0.01&12.53&0.013$-$0.026&0.073$-$0.089&10.44$-$10.97 & \\%
\hline

$H_{s}$&4&0.001&0.1&39.36&0.003$-$0.004&0.005$-$0.007&197.60$-$287.79 & \\%
$H_{b_1}$&4&0.001&0.1&39.36&0.014$-$0.018&0.027$-$0.032&43.65$-$46.53 & 2.09\\%
$H_{b_2}$&4&0.001&0.1&39.36&0.014$-$0.018&0.039$-$0.041&27.79$-$25.73 & \\%
\end{tabular}%
\label{tab:m4}
\end{table*}

\begin{table*}[htpb!]
\caption{Solutions of the dispersion relation for $m=12$ and $\gamma=12$}
\begin{tabular}{c|cccccccc}%
Mode&m&$\varepsilon_\mathrm{j}$&$\eta$&$M_\mathrm{j}$&$\omega$ [$c/R_\mathrm{j}$]&$\Re(k)$ [$1/R_\mathrm{j}$]&
$\lambda$ [mas] & Dispersion\\%
\hline%
\hline%
$H_{s}$&12&0.01&$10^{-5}$&4.21&0.0012$-$0.0036&0.0034$-$0.0083&133.23$-$257.61 & \\%
$H_{b_1}$&12&0.01&$10^{-5}$&4.21&0.0039$-$0.0079&0.0230$-$0.0290&31.15$-$34.43 & 0.912\\%
$H_{b_2}$&12&0.01&$10^{-5}$&4.21&0.0045$-$0.0088&0.0370$-$0.0430&19.24$-$20.25 & \\%
\hline

$H_{s}$&12&0.01&$6 \times 10^{-5}$&4.21&0.0030$-$0.0080&0.0077$-$0.0185&59.55$-$117.70 & \\%
$H_{b_1}$&12&0.01&$6 \times 10^{-5}$&4.21&0.0080$-$0.0150&0.0455$-$0.0547&16.56$-$17.54 & 0.280\\%
$H_{b_2}$&12&0.01&$6 \times 10^{-5}$&4.21&0.0120$-$0.0240&0.0700$-$0.0850&10.78$-$11.34 & \\%
\hline

$H_{s}$&12&0.1&$10^{-4}$&4.21&0.0030$-$0.010&0.0090$-$0.0230&47.94$-$93.90 & \\%
$H_{b_1}$&12&0.1&$10^{-4}$&4.21&0.0110$-$0.0210&0.0530$-$0.0650&14.94$-$15.66 & 0.376\\%
$H_{b_2}$&12&0.1&$10^{-4}$&4.21&0.0140$-$0.0280&0.0840$-$0.1020&8.89$-$9.40 & \\%
\end{tabular}%
\label{tab:m12}
\end{table*}

\begin{table*}[htpb!]
\caption{Solutions of the dispersion relation for $m=16$ and $\gamma=5$}
\begin{tabular}{c|cccccccc}%
Mode&m&$\varepsilon_\mathrm{j}$&$\eta$&$M_\mathrm{j}$&$\omega$ [$c/R_\mathrm{j}$]&$\Re(k)$ [$1/R_\mathrm{j}$]&
$\lambda$ [mas] & Dispersion\\%
\hline%
\hline%
$H_{s}$&16&0.01&0.001&12.53&0.004$-$0.010&0.012$-$0.027&37.26$-$67.77 & \\%
$H_{b_1}$&16&0.01&0.001&12.53&0.019$-$0.036&0.062$-$0.084&13.68$-$15.38 & 0.508\\%
$H_{b_2}$&16&0.01&0.001&12.53&0.014$-$0.026&0.103$-$0.118&7.10$-$7.36 & \\%
\hline
$H_{s}$&16&0.01&0.002&12.53&0.004$-$0.011&0.012$-$0.028&35.49$-$68.72 & \\%
$H_{b_1}$&16&0.01&0.002&12.53&0.021$-$0.037&0.071$-$0.094&11.55$-$13.20 & 0.590\\%
$H_{b_2}$&16&0.01&0.002&12.53&0.025$-$0.054&0.123$-$0.156&6.45$-$6.71 & \\%
\hline
$H_{s}$&16&0.1&$2\times 10^{-5}$&4.21&0.003$-$0.008&0.010$-$0.021&50.50$-$85.77 & \\%
$H_{b_1}$&16&0.1&$2\times 10^{-5}$&4.21&0.008$-$0.016&0.056$-$0.068&12.64$-$13.86 & 0.467\\%
$H_{b_2}$&16&0.1&$2\times 10^{-5}$&4.21&0.009$-$0.017&0.088$-$0.100&7.93$-$8.32 & \\%
\end{tabular}%
\label{tab:W5}
\end{table*}

\begin{table*}[htpb!]
\caption{Solutions of the dispersion relation for $\gamma=17$}
\begin{tabular}{c|cccccccc}%
Mode&m&$\varepsilon_\mathrm{j}$&$\eta$&$M_\mathrm{j}$&$\omega$ [$c/R_\mathrm{j}$]&$\Re(k)$ [$1/R_\mathrm{j}$]&
$\lambda$ [mas] & Dispersion\\%
\hline%
\hline%
$H_{s}$&25&0.01&0.001&12.55&0.002$-$0.005&0.005$-$0.010&116.90$-$206.14&\\%
$H_{b_1}$&25&0.01&0.001&12.55&0.009$-$0.015&0.028$-$0.036&30.57$-$34.39&0.380\\%
$H_{b_2}$&25&0.01&0.001&12.55&0.017$-$0.026&0.050$-$0.062&18.50$-$19.57&\\%
\hline
$H_{s}$&25&0.01&0.002&12.55&0.003$-$0.006&0.005$-$0.011&115.94$-$195.36&\\%
$H_{b_1}$&25&0.01&0.002&12.55&0.011$-$0.018&0.031$-$0.040&29.61$-$32.61&0.310\\%
$H_{b_2}$&25&0.01&0.002&12.55&0.017$-$0.027&0.052$-$0.064&17.66$-$18.55&\\%
\hline
$H_{s}$&25&0.1&$2\times 10^{-5}$&4.21&0.004$-$0.009&0.010$-$0.020&56.13$-$89.37&\\%
$H_{b_1}$&25&0.1&$2\times 10^{-5}$&4.21&0.011$-$0.021&0.040$-$0.052&21.56$-$22.45&0.165\\%
$H_{b_2}$&25&0.1&$2\times 10^{-5}$&4.21&0.012$-$0.022&0.063$-$0.077&11.87$-$12.76&\\%
\hline%
\hline%
$H_{s}$&16&0.01&0.001&12.53&0.002$-$0.004&0.004$-$0.009&125.88$-$219.63 & \\%
$H_{b_1}$&16&0.01&0.001&12.53&0.008$-$0.014&0.027$-$0.035&32.00$-$35.60 & 0.531\\%
$H_{b_2}$&16&0.01&0.001&12.53&0.014$-$0.024&0.047$-$0.058&19.03$-$19.93 & \\%
\hline
$H_{s}$&16&0.01&0.002&12.53&0.003$-$0.005&0.005$-$0.010&122.05$-$203.81 & \\%
$H_{b_1}$&16&0.01&0.002&12.53&0.010$-$0.016&0.030$-$0.038&29.80$-$32.57 & 0.400\\%
$H_{b_2}$&16&0.01&0.002&12.53&0.014$-$0.024&0.050$-$0.061&17.67$-$18.51 & \\%

$H_{s}$&16&0.01&0.01&12.55&0.003$-$0.005&0.005$-$0.010&122.05$-$203.81&\\%
$H_{b_1}$&16&0.01&0.01&12.55&0.010$-$0.016&0.030$-$0.038&29.80$-$32.57&0.397\\%
$H_{b_2}$&16&0.01&0.01&12.55&0.014$-$0.024&0.050$-$0.061&17.67$-$18.51&\\%
\hline
$H_{s}$&16&0.1&$10^{-4}$&4.21&0.004$-$0.010&0.010$-$0.023&49.14$-$90.37&\\%
$H_{b_1}$&16&0.1&$10^{-4}$&4.21&0.015$-$0.029&0.051$-$0.067&17.27$-$18.58& 0.319\\%
$H_{b_2}$&16&0.1&$10^{-4}$&4.21&0.015$-$0.027&0.086$-$0.103&8.74$-$9.25&\\%
\hline
$H_{s}$&16&0.1&$2\times 10^{-5}$&4.21&0.002$-$0.006&0.006$-$0.013&84.23$-$144.92 & \\%
$H_{b_1}$&16&0.1&$2\times 10^{-5}$&4.21&0.006$-$0.010&0.032$-$0.038&24.15$-$25.40 & 0.088\\%
$H_{b_2}$&16&0.1&$2\times 10^{-5}$&4.21&0.009$-$0.018&0.048$-$0.059&16.03$-$16.99 & \\%
\end{tabular}%
\label{tab:W17}
\end{table*}

\begin{figure*}[htpb!]
\centering
\includegraphics[width=0.45\textwidth]{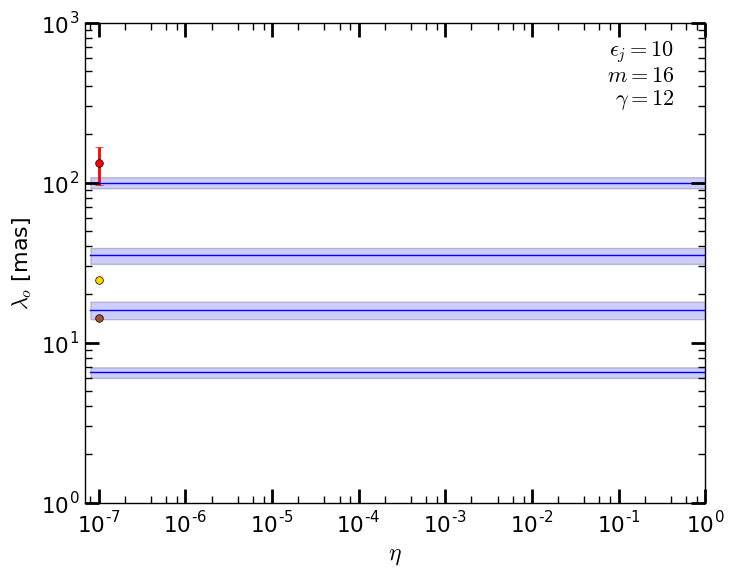}
\includegraphics[width=0.45\textwidth]{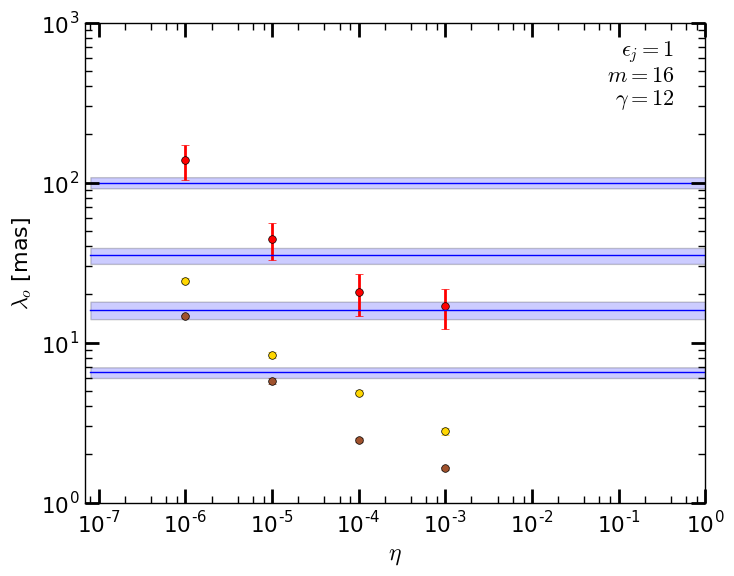}
\includegraphics[width=0.45\textwidth]{epsi01m16.png}
\includegraphics[width=0.45\textwidth]{epsi001m16.png}
\includegraphics[width=0.45\textwidth]{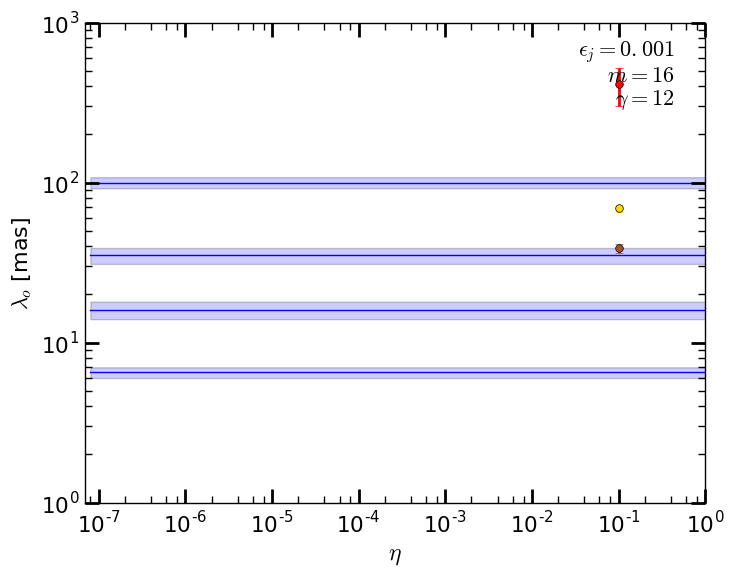}
\caption{Corresponding observed wavelengths from the calculations for $m=16$. The color scale and the meaning of the horizontal lines are the same as used in Fig~\ref{fig:simresults}. The corresponding jet internal energy for each plot is written in the figure.}
\end{figure*}

\begin{figure*}[htpb!]
\centering
\includegraphics[width=0.45\textwidth]{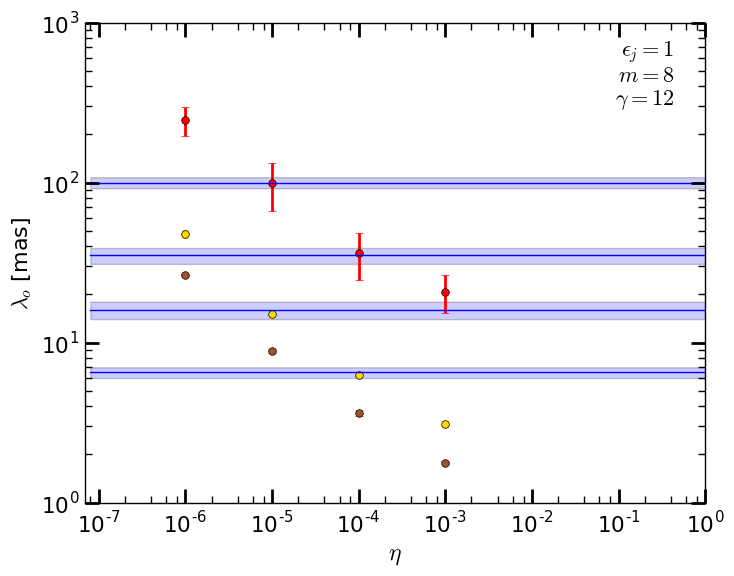}
\includegraphics[width=0.45\textwidth]{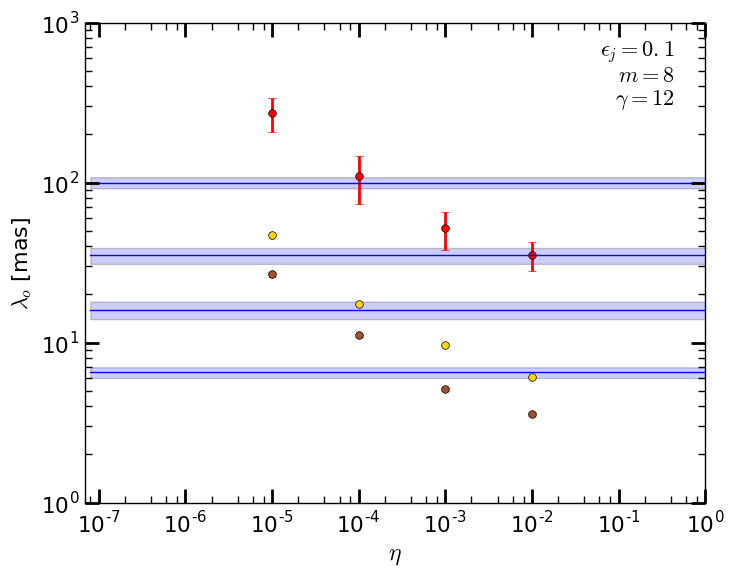}
\includegraphics[width=0.45\textwidth]{epsi001m8.png}
\includegraphics[width=0.45\textwidth]{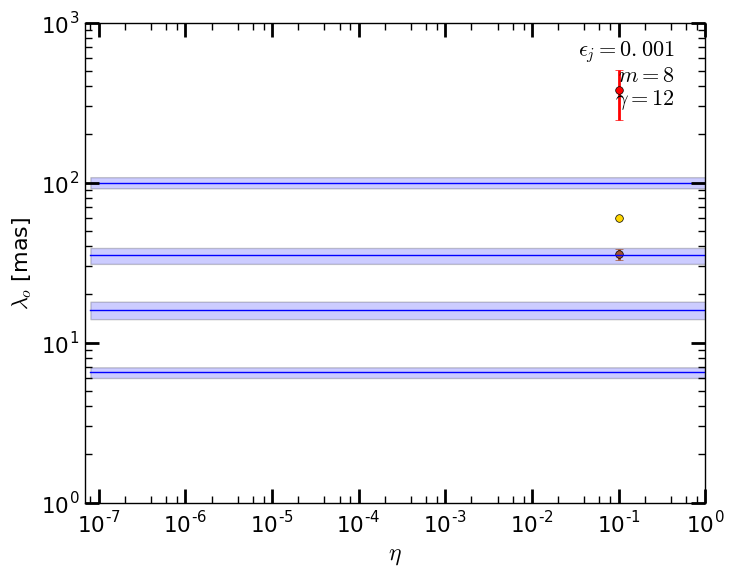}
\caption{Corresponding observed wavelengths from the calculations for $m=8$. The color scale and the meaning of the horizontal lines are the same as used in Fig~\ref{fig:simresults}. The corresponding jet internal energy for each plot is written in the figure.}
\end{figure*}

\begin{figure*}[htpb!]
\centering
\includegraphics[width=0.45\textwidth]{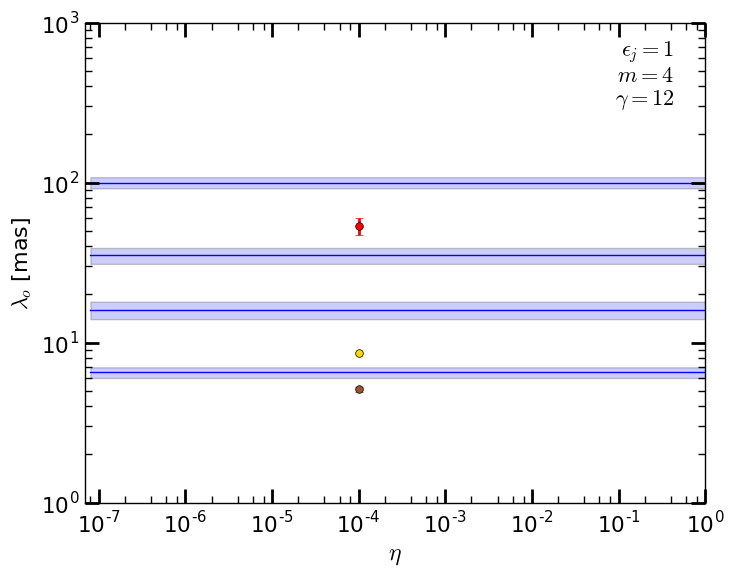}
\includegraphics[width=0.45\textwidth]{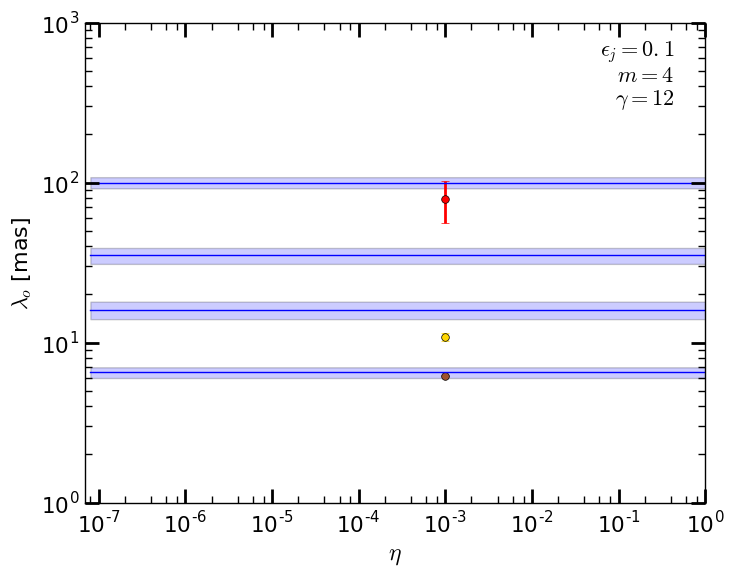}
\includegraphics[width=0.45\textwidth]{epsi001m4.png}
\includegraphics[width=0.45\textwidth]{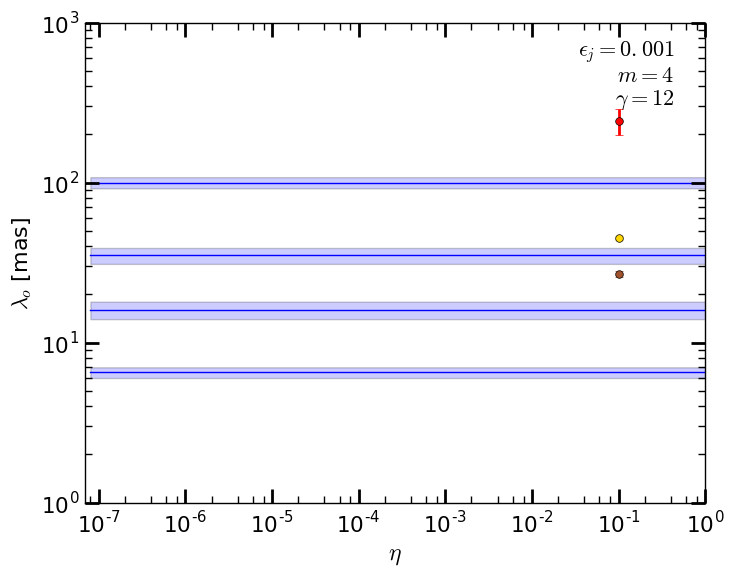}
\caption{Corresponding observed wavelengths from the calculations for $m=4$. The color scale and the meaning of the horizontal lines are the same as used in Fig~\ref{fig:simresults}. The corresponding jet internal energy for each plot is written in the figure.}
\end{figure*}

\begin{figure*}[htpb!]
\centering
\includegraphics[width=0.45\textwidth]{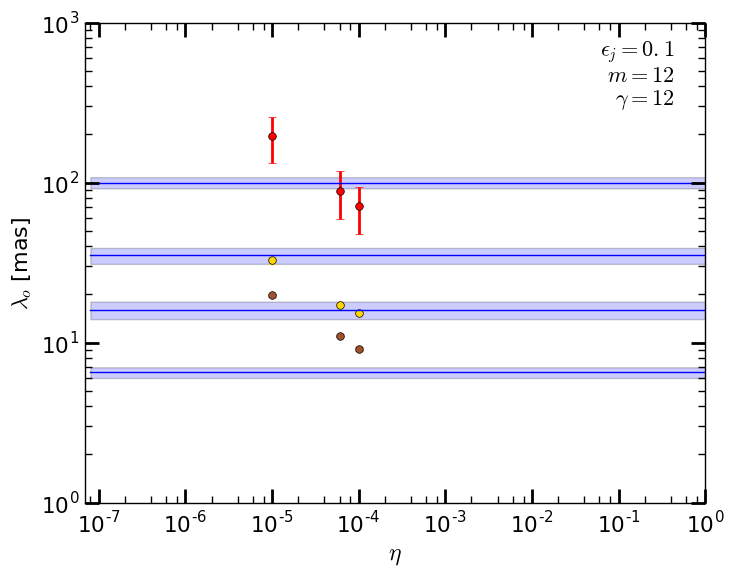}
\caption{Corresponding observed wavelengths from the calculations for $m=12$. The color scale and the meaning of the horizontal lines are the same as used in Fig~\ref{fig:simresults}. The corresponding jet internal energy for each plot is written in the figure.}
\end{figure*}

\begin{figure*}[htpb!]
\centering
\includegraphics[width=0.45\textwidth]{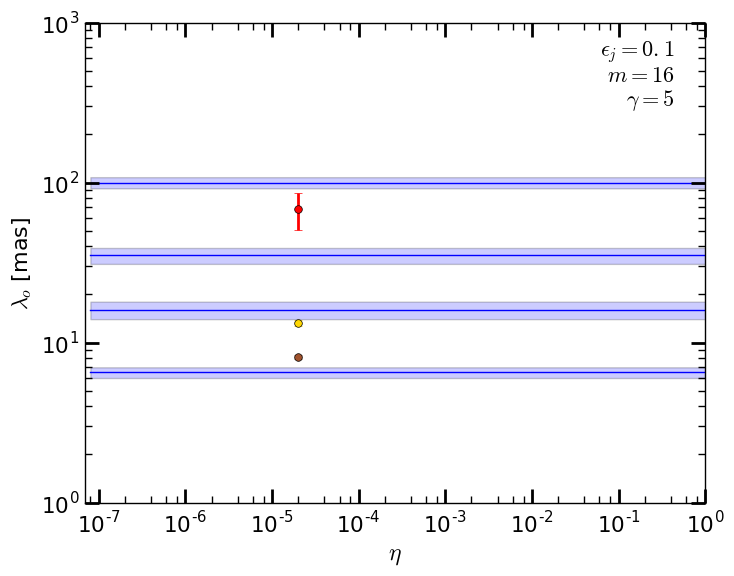}
\includegraphics[width=0.45\textwidth]{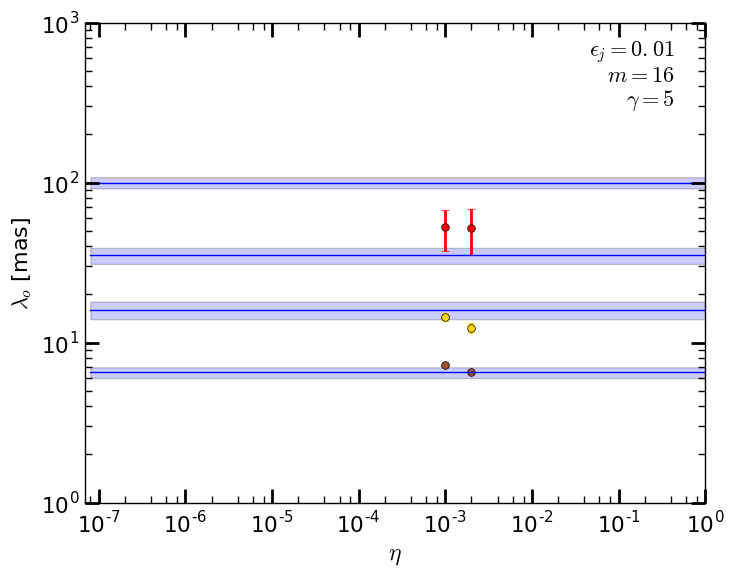}
\caption{Corresponding observed wavelengths from the calculations for $m=16$ and $\gamma = 5$. The color scale and the meaning of the horizontal lines are the same as used in Fig~\ref{fig:simresults}. The corresponding jet internal energy for each plot is written in the figure.}
\end{figure*}

\begin{figure*}[htpb!]
\centering
\includegraphics[width=0.45\textwidth]{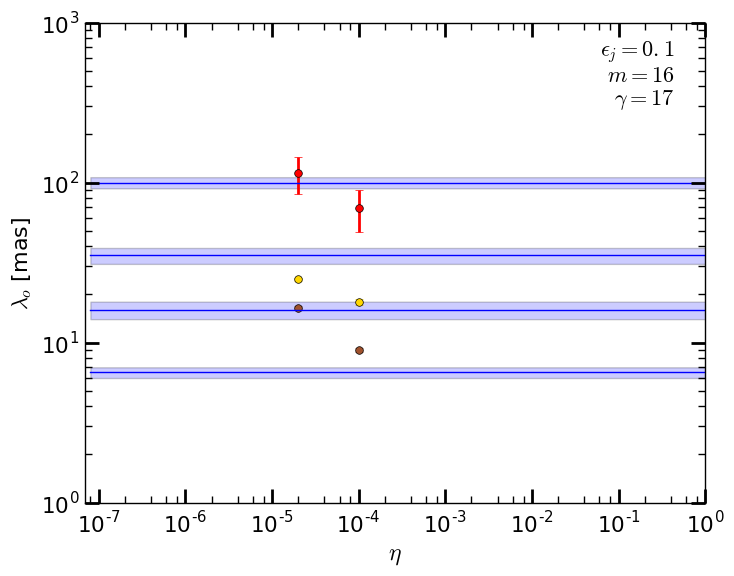}
\includegraphics[width=0.45\textwidth]{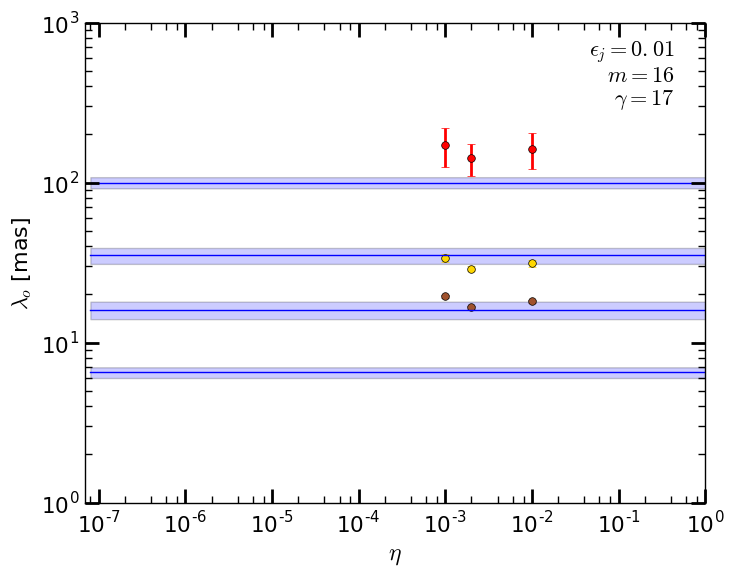}
\caption{Corresponding observed wavelengths from the calculations for $m=16$ and $\gamma = 17$. The color scale and the meaning of the horizontal lines are the same as used in Fig~\ref{fig:simresults}. The corresponding jet internal energy for each plot is written in the figure.}
\end{figure*}

\begin{figure*}[htpb!]
\centering
\includegraphics[width=0.45\textwidth]{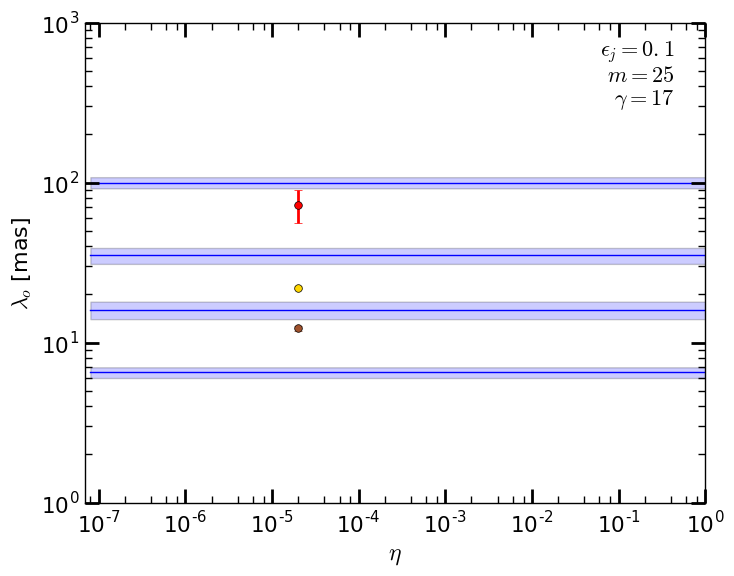}
\includegraphics[width=0.45\textwidth]{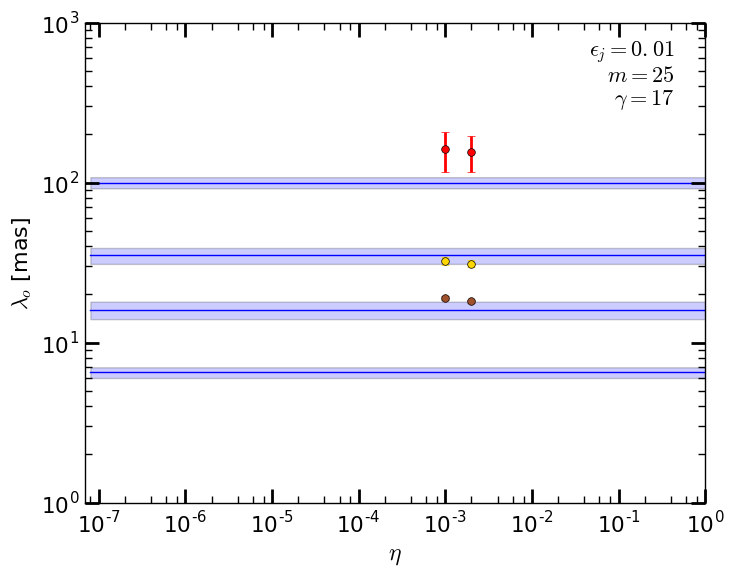}
\caption{Corresponding observed wavelengths from the calculations for $m=25$ and $\gamma = 17$. The color scale and the meaning of the horizontal lines are the same as used in Fig~\ref{fig:simresults}. The corresponding jet internal energy for each plot is written in the figure.}
\end{figure*}

\end{appendix}

\end{document}